\documentclass[aps,prb,twocolumn,amsmath,amssymb,nofootinbib,eqsecnum,tighten,showpacs,floatfix]{revtex4}

\usepackage{graphicx}
\usepackage{dcolumn} 
\usepackage{bm}      
\usepackage{epic}

\begin{document}

\title{Zero-modes in the random hopping model}

\author{P.\ W.\ Brouwer}
\affiliation{Laboratory of Atomic and Solid State Physics,
             Cornell University, 
             Ithaca, 
             NY 14853-2501,
             USA}
\author{A.\ Furusaki}
\affiliation{Yukawa Institute for Theoretical Physics,
             Kyoto University,
             Kyoto 606-8502, 
             Japan}
\author{Y.\ Hatsugai}
\affiliation{Department of Applied Physics,
             University of Tokyo,
             7-3-1 Hongo Bunkyo-ku,
             Tokyo 113-8656,
             Japan}
\affiliation{PRESTO, 
             Japan Science and Technology Corporation,
             Kawaguchi-shi, 
             Saitama, 
             332-0012, 
             Japan}
\author{Y.\ Morita}
\affiliation{Department of Applied Physics,
             University of Tokyo,
             7-3-1 Hongo Bunkyo-ku,
             Tokyo 113-8656,
             Japan}
\author{C.\ Mudry}
\affiliation{Paul Scherrer Institute,
             CH-5232 Villigen PSI, 
             Switzerland}
\affiliation{Yukawa Institute for Theoretical Physics,
             Kyoto University,
             Kyoto 606-8502, 
             Japan}
\author{E.\ Racine}
\affiliation{Laboratory of Atomic and Solid State Physics, 
             Cornell University, 
             Ithaca, 
             NY 14853-2501,
             USA}

\date{\today}

\begin{abstract}
If the number of lattice sites is odd, a quantum particle hopping
on a bipartite lattice with random hopping 
between the two sublattices only is guaranteed to have an
eigenstate at zero energy. We show that the localization length
of this eigenstate depends strongly on the boundaries of the lattice,
and can take values anywhere between the mean free path and infinity. 
The same dependence on boundary conditions is seen in the conductance
of such a lattice if it is connected to electron reservoirs via
narrow leads. For any nonzero energy, the dependence on boundary 
conditions is removed for sufficiently large system sizes.
\end{abstract}


\maketitle

%
%
%
%
%
%
%
%
%
%
%

\section{Introduction}

Zero-modes, wavefunctions at zero energy, often arise in problems
when a quantum particle moves in a background with a nontrivial 
topological structure.\cite{Callan85}$^{-}$\cite{Altland99} 
Quantum fluctuations associated to these zero-modes 
have dramatic physical consequences. They appear both in field
theories, related to chiral and parity anomalies, \cite{Callan85}
and in lattice regularization of field theories,\cite{Janssen96}
and have applications to a wide range of areas in physics: 
Chiral symmetry breaking in 
(1+1)--space-time quantum electrodynamics,\cite{Schwinger62}
edge states along the boundary of a disk threaded by a magnetic 
flux,\cite{Aharonov79,Moroz95} 
singular contributions to the Hall 
conductance from electrons hopping on a square lattice in the presence of
a uniform magnetic 
field,\cite{Kohmoto89}
superconductivity of a cosmic 
string,\cite{Witten85}
localization of a fractional charge at a domain wall in a 
charge density wave,\cite{Jackiw76,Su79,Su81,Takayama80,Golstone86,Su86}
induction of a persistent mass current in 
$^{3}$He-A,\cite{Stone85,Volovik86,Gaitan87}
antiphase boundaries in narrow-gap semiconductors,\cite{Fradkin86}
surface\cite{Hu94} (edge\cite{Volovik97,Laughlin98}) states 
in a superconductor with $d_{x^2-y^2}(+id_{xy})$ symmetry
or in a chiral $p$-wave\cite{Buchholtz81} superconductor,
edge states in nanographite ribbon junctions,\cite{Wakabayashi00}
and
itinerant-electron ferromagnetism in the repulsive Hubbard model.\cite{Lieb89}

Related nontrivial topological structures can also exist
in random matrix theory\cite{RMT}
and
in the problem of Anderson localization when the disorder possesses
a special symmetry. 
In this context, an almost half a century old example 
is that of a one-dimensional chain with link disorder.\cite{Dyson53} 
Here, zero energy corresponds to the center of the energy band, 
and an eigenfunction with zero energy is guaranteed to exist if the number 
of sites in the chain is odd.\cite{Fleishman77}
In the thermodynamic limit,
the density of states\cite{Dyson53} 
and the localization length\cite{Theodorou76,Eggarter78}
are singular at zero energy, whereas correlation functions
of the local density of states\cite{Shelton98,Hastings01} 
are algebraic functions.
This anomalous behavior is an example of a 
strongly random critical point.\cite{Fisher94}
Its origin is rooted in the stochastic properties
of the zero-modes supported by the Dirac equation in one-space dimension
in the background of a
white-noise correlated random mass.\cite{Shelton98,Hastings01,Balents97}
More recently, a two-dimensional random Dirac Hamiltonian with
white-noise correlated $U(1)\times SU(N)$ 
random vector potential was shown to be
critical at zero energy.\cite{Ludwig94,Nersesyan94} 
As with the stochastic model of Dyson, this critical behavior  
can be ascribed to the stochastic properties of zero-modes 
supported by Dirac equations in two-space dimensions
in the background of white noise correlated 
random vector potentials.\cite{Mudry96,Chamon96,Kogan96}

\begin{figure}

\includegraphics[width=\columnwidth]{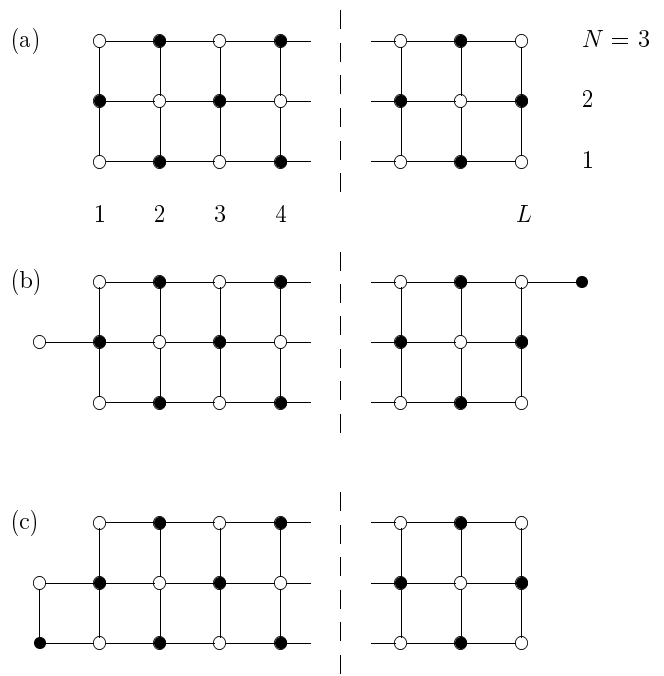}

\caption{\label{fig: 3 exs of lattice with neq BC} 
Three examples of a lattice with different
boundary conditions. The sublattices $A$ and $B$ correspond to
the white and black sites, respectively. In all three examples,
the number $N_A$ of sites on sublattice $A$ is one more than
the number $N_B$ of sites on sublattice $B$. 
(a) 
Conventional rectangular shaped wire with $N=3$ and odd $L\gg N$.
(b) 
Boundary conditions with respect to (a)
have been changed by adding a white site
to the left and adding a black site to the right. 
(c) 
Boundary conditions with respect to (a)
have been changed by adding a pair of white and black sites to the left.}

\end{figure}

While the critical behavior of the (continuum) Dirac equation is
related to zero-modes in the infinite system, the existence of
zero-modes for Dyson's stochastic model of a quantum particle hopping
on a bipartite lattice with link disorder can also occur for a finite
system size. The Hamiltonian for this system is
\begin{eqnarray}
  {\mathcal{H}} = 
  \sum_{i,j} 
  t^{\vphantom{\dagger}}_{ij} 
  c^{\dagger}_i c^{\vphantom{\dagger}}_j,
\label{eq:Schr Eq in intro}
\end{eqnarray}
where $i$ and $j$ label the lattice sites on a cartesian grid in $d$ 
dimensions, say, and the hopping matrix elements $t_{ij}$ are nonzero for 
nearest-neighbors only. Examples of bipartite lattices are depicted
in Fig.\ \ref{fig: 3 exs of lattice with neq BC}. 
In general, the $t_{ij}$ will have a small
random component in addition to an average $t$ which sets the
width of the spectrum of ${\mathcal{H}}$. In this paper we refer to 
Eq.\ (\ref{eq:Schr Eq in intro}) with the random $t_{ij}$ as the
``random hopping model''.
The special case when it is only the phase of the hopping amplitude 
$t_{ij}$ that is random
is also known as the random flux problem.

For the Hamiltonian (\ref{eq:Schr Eq in intro}), 
the existence of the zero-modes follows from the 
existence of a ``sublattice'' (or ``chiral'') symmetry. This symmetry 
follows when the lattice is divided into two 
sublattices $A$ and $B$ such that the hopping matrix $t_{ij}$ only connects 
sites from the two sublattices, but not sites of the same sublattice.
For the example of Fig.\ \ref{fig: 3 exs of lattice with neq BC}, 
the sublattices $A$ and $B$ 
correspond to the white and black sites, respectively.
In a matrix form and after a relabeling of indices, 
the eigenvalue problem ${\mathcal{H}} |\psi\rangle = \varepsilon
|\psi\rangle$ can be rewritten as
\begin{eqnarray}
&&
\varepsilon 
\left(\begin{array}{cc}
{\psi}^{\   }_A
\\
{\psi}^{\   }_B
\end{array}\right)
=
\left(\begin{array}{cc}
0 
&
{t}^{\   }_{AB}
\\
{t}^{\dag}_{AB} 
&
0
\end{array}\right)
\left(\begin{array}{cc}
{\psi}^{\   }_A
\\
{\psi}^{\   }_B
\end{array}\right),
\label{eq:matrix H in sublattice grading}
\end{eqnarray}
where $\psi_A$ and $\psi_B$ denote the wavefunction on the lattice
sites of the sublattices $A$ and $B$, respectively. 
Then, denoting the number of sites in the
sublattices $A$ and $B$ by $N_A$ and $N_B$,
counting dimensions
in Eq.\ (\ref{eq:matrix H in sublattice grading}) immediately yields that  
the number of linearly independent zero-modes is 
$|N_{A} - N_{B}|$.\cite{Sutherland86}
To see this, note that if $N_A > N_B$,
$\psi_A$ obeys an underdetermined set of linear equations, 
while $\psi_B$ obeys an overdetermined set of equations.
(For all lattices shown in Fig.\ \ref{fig: 3 exs of lattice with neq BC}, 
there is one zero-mode with support on sublattice $A$.)

Addition or removal of a single site changes the
number of zero-modes, as it changes the difference $|N_A - N_B|$
between the numbers of sites in the two sublattices. 
At the same time, the singular behavior of the density of states
and the localization length in the random hopping model in the thermodynamic
limit are considered ``intrinsic'' properties, i.e., 
they are derived from continuum models and should not depend
on boundary conditions of the lattice. 
Hence, whilst both the existence of zero-modes for lattices with boundaries
and the singular behavior of the localization length are
manifestations of the same sublattice symmetry, they are so in
very different ways. One might even ask to what extent the localization
length of zero-modes is representative for the ``intrinsic'' 
localization length of the random hopping problem or random flux problem 
on a lattice without boundaries. 
This is the question addressed in this paper. 

Our answer is that the localization length of the zero-modes for 
lattices with boundaries is not an ``intrinsic'' property of 
the random hopping model. 
After a brief review of the transfer matrix formalism
in Subsec.\ \ref{subsec:Transfer matrix formalism},
we support this conclusion 
in Subsec.\ \ref{subsec:Zero energy}
by analytical and numerical solution of the problem
in the case of a wire geometry: Depending on the boundary 
conditions, zero-modes exhibit a range of localization lengths, the
smallest one being of the order of the mean free path.
In Subsec.\ \ref{subsec:Nonzero energy} 
we then show that this extreme sensitivity to
boundary conditions is an anomaly corresponding to the special
case $\varepsilon=0$. For {\em any} energy $\varepsilon \neq 0$ there 
is a unique localization
length if the system size is sufficiently large. For sufficiently
small $\varepsilon$, this unique
localization length coincides with the largest of the possible
localization lengths at $\varepsilon=0$. 
We discuss higher dimensional examples in Sec.\ 
\ref{sec:Higher dimensional examples} and conclude in Sec.\
\ref{sec:Conclusions}.

\section{Quasi-one-dimensional geometry}
\label{sec:Quasi-one-dimensional geometry}

In this section we consider a two-dimensional lattice, 
$N$ sites wide and $L \gg N$ sites long. 
Examples for $N=3$ are given in 
Fig.\ \ref{fig: 3 exs of lattice with neq BC}
whereby sites on sublattices $A$ and $B$
are colored in white and black, respectively. 
The number of zero-modes equals $|N_A- N_B|$. 
For definiteness, we assume $N_A > N_B$, so that
all zero-modes have support on sublattice $A$. 

In this section, we will compare zero-modes in wire geometries 
with different boundary conditions at the two ends of the wire, 
as is shown, e.g., in Fig.\ \ref{fig: 3 exs of lattice with neq BC}.
We will establish that the localization length of these zero-modes
cannot be thought of as being intrinsic, i.e., independent of the
boundary conditions even as the thermodynamic limit $L\to\infty$ is taken.
We will then turn our attention to finite energies and, supported by a 
numerical solution of the problem, argue that
an intrinsic localization length at arbitrarily small but finite
energies does indeed exist. The order in which the limits $L\to\infty$
and $\varepsilon\to0$ are taken is thus essential for the extraction of
an intrinsic localization length at zero energy.

The sublattice symmetry singles out the band center in that,
under an appropriate choice of boundary conditions,
an exact energy eigenfunction at that energy can be constructed for
any realization of the disorder. This is not true of any finite
energy $\varepsilon$ in a closed and finite system.
Therefore we will proceed in two steps. We first compare the
localization length of zero-modes in a closed system with
the exponential decay lengths for transmission probability of
plane waves in an open system and establish that they are
equal. Then we study how the transmission probability of
plane waves is changed when the energy becomes finite. In both steps
we use the transfer matrix formalism, which can deal with open and
closed systems in a unified way.

\subsection{Transfer matrix formalism in an unbounded wire}
\label{subsec:Transfer matrix formalism}

\begin{figure}

\includegraphics[width=\columnwidth]{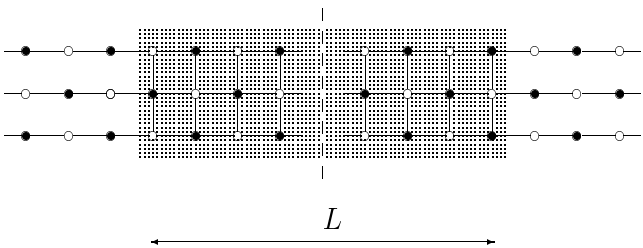}

\caption{\label{fig: conventional wire coupled to leads} 
Disordered quantum wire of even length $L$ (dotted region)
coupled to left and right reservoirs 
through leads of the same width as the wire.
We have chosen leads without transverse hopping for
technical convenience.}

\end{figure}

\subsubsection{Plane-wave representation}

In the absence of disorder, the eigenfunctions of the Schr\"odinger
equation (\ref{eq:Schr Eq in intro}) in a wire geometry 
as depicted in Fig.\ \ref{fig: conventional wire coupled to leads} are plane
waves. At zero energy, there are $N$ independent wavefunctions for
plane waves traveling to the right, and $N$ independent wavefunctions
traveling to the left. An arbitrary wavefunction can be expanded in
the basis of plane waves. 
In the presence of disorder,
the plane-wave expansion coefficients $a_{n}$, $n=1,\ldots,2N$ acquire 
a dependence on the position $y$ along 
the wire. The relation between the $a_{n}(y)$ at different positions
along the wire can be expressed through the transfer matrix ${\cal M}$,
\begin{equation}
  a_{n}(y) = \sum_{m=1}^{2N} {\cal M}_{nm}(y,y')a_{m}(y').
  \label{eq:plane wave trs matrix}
\end{equation}
Current conservation and the sublattice symmetry imply that
${\cal M}$ can be parametrized as\cite{Brouwer98}
\begin{eqnarray}
{\mathcal{M}}=
\left(\begin{array}{cc}
U
&
0
\\
&
\\
0
&
U
\end{array}\right)
\left(\begin{array}{cc}
\cosh X
&
\sinh X
\\
&
\\
\sinh X
&
\cosh X
\end{array}\right)
\left(\begin{array}{cc}
V
&
0
\\
&
\\
0
&
V
\end{array}\right).
\label{eq:polar decomposition trsf matrix} \label{eq:Mxxp}
\end{eqnarray}
The $2\times2$ grading displayed here is that of right and left moving
plane waves. The $N\times N$ matrices $U$ and $V$ are
unitary ($\beta=2$), symplectic ($\beta=4$), and orthogonal ($\beta=1$)
when the hopping amplitudes are complex, real quaternions, and real valued, 
respectively. 
The $N\times N$ matrix $X = \mbox{diag}(x_1,\ldots,x_N)$ 
is real valued and diagonal. 
In the absence of the sublattice symmetry, the transfer matrix 
has a similar parametrization.\cite{Beenakker97}
The main difference between the
cases with and without sublattice symmetry is that without sublattice
symmetry the $x_n$ can always be chosen positive, while with
sublattice symmetry both positive and negative $x_n$ appear.

The localization properties of the disordered wire are encoded in 
the transfer matrix ${\mathcal{M}}$. They are dominated by the
diagonal matrix $X$ in Eq.\ (\ref{eq:Mxxp}).
The distribution of the $x_n$ at zero energy has been studied
in Ref.\ \onlinecite{Brouwer98} for the case of a wire with random 
hopping only. In terms of the mean free path $l$ of the wire, it
was found that
\begin{equation}
  x_n = \frac{L}{\xi_n},
  \quad \ n=1,\ldots,N,
\label{eq:averaged x_n}
\end{equation}
if $L \gg N l$, up to fluctuations of relative order 
$(N l/L)^{-1/2}$, where
the Lyapunov exponents $1/\xi_n$ (the inverse localization
lengths) are given by (to leading and subleading order in $N$,
see Ref.\ \onlinecite{Brouwer00})
\begin{equation}
  \frac{1}{\xi_n} = \frac{\beta(N+1-2n)}{(\beta N + 2 - \beta)l},
  \qquad
  n=1,\ldots,N. 
  \label{eq:xin}
\end{equation}

For an infinite wire, the conductance is given by
\begin{equation}
  G = 
  \frac{2e^2}{h} \sum_{n=1}^{N} \cosh^{-2} x_n\equiv
  \frac{2e^2}{h} g.
\end{equation}
Hence, for an infinite
wire it is only the smallest in magnitude of the Lyapunov
exponents that governs the exponential decay of the conductance. 
Its inverse is thus identified with the localization
length $\xi$ of the system,
\begin{equation}
  \xi = 
  \left\{ 
  \begin{array}{cc}
  (\beta N + 2 - \beta)l/\beta
  & 
  \mbox{for $N$ even},\\
  & \\
  \infty 
  & 
  \mbox{for $N$ odd}. 
  \end{array} 
  \right.
\end{equation}
As we will find below
that all localization lengths $\xi_n$ can serve as localization
lengths for zero-modes in a finite sized wire, we will refer to 
their maximum $\xi$ as the localization length for an infinite wire.

\begin{figure}

\centering
\includegraphics[width=0.7\columnwidth]{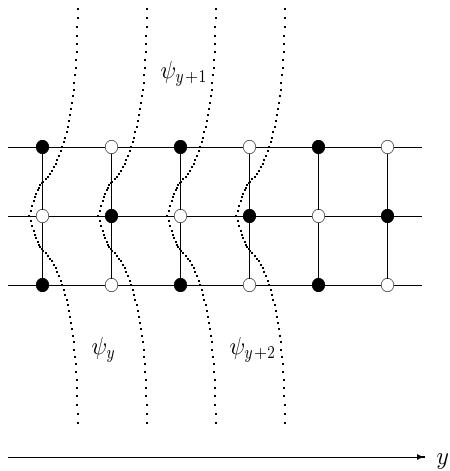}

\caption{\label{fig:choice} 
Choice of the vectors $\psi_y$ of Eq.\
(\protect\ref{eq:Schr Eq for transfer matrix}).}
\end{figure}

\subsubsection{Site representation}

An alternative representation for the transfer matrix is obtained
using a site representation for the wavefunction instead of an 
expansion in plane waves. The sublattice symmetry becomes manifest
in the site representation if the wavefunction elements are arranged
in $N$-component vectors $\psi_y$ containing elements of one
sublattice only. In that notation, the Schr\"odinger equation
(\ref{eq:Schr Eq in intro}) reads
\begin{eqnarray}
  \varepsilon \psi^{\vphantom{\dag}}_y=
  T^{\vphantom{\dag}}_y            \psi^{\vphantom{\dag}}_{y+1}
 +T^{          \dag }_{y-1}        \psi^{\vphantom{\dag}}_{y-1}.
\label{eq:Schr Eq for transfer matrix}
\end{eqnarray}
The index $y$ labels the coordinate along the wire and the
$N$-component vectors $\psi_y$ contain wavefunction elements
on sites of sublattice $A$ if $y$ is odd and of sublattice $B$ if
$y$ is even. A possible
choice for the vectors $\psi_y$ is shown in Fig.\ \ref{fig:choice}. 
The $N\times N$ matrices $T_{y}$ contain the hopping amplitudes 
between adjacent sites.

The solution of Eq.\ (\ref{eq:Schr Eq for transfer matrix}) can be
represented in terms of a transfer matrix as well,
\begin{eqnarray}
\left(\begin{array}{cc}
\psi_{y  }
\\
\psi_{y-1}
\end{array}\right) &=&
{\mathcal{M}}(y,y')
\left(\begin{array}{cc}
\psi_{y'}
\\
\psi_{y'-1}
\end{array}\right),
\nonumber \\
{\mathcal{M}}(y,y') &=&
\prod_{m=y'}^{y-1}
  \left[ \left( \begin{array}{cc}
  T_m^{-1} & 0 \\ 0 & 1 \end{array} \right)
  \left( \begin{array}{cc}
  \varepsilon & -T_{m-1}^{\dag} \\
  1 & 0 \end{array} \right) \right].
\nonumber \\&&
  \label{eq:Wannier}
\end{eqnarray}
We note that, if $y-y'>0$ is even, the transfer matrix ${\cal M}(y,y')$ 
is block diagonal at zero energy. In that case, ${\cal M}(y,y')$ can be
parametrized as
\begin{eqnarray}
  {\cal M}(y,y') &=&
  \left(\begin{array}{cc} 
  U 
  e^X
  V & 
  0 \\ 
  0 & 
  T_{y-1}^{-1\dagger} U
  e^{-X}
  V T_{y'-1}^{\dagger}
  \end{array} \right).
  \label{eq:Mxxs}
\end{eqnarray}
Here $U$ and $V$ are orthogonal (unitary, symplectic) matrices
for $\beta=1$ ($\beta=2,4$) and $X$ is a diagonal matrix.
The grading used in Eq.\ (\ref{eq:Mxxs}) corresponds to the division
into sublattices $A$ and $B$.
As the transfer matrices of Eqs.\ (\ref{eq:Wannier}) and 
(\ref{eq:plane wave trs matrix}) are related by
a simple basis transformation, the distribution of the eigenvalues 
$x_n$ of the matrix $X$ is also given by 
Eqs.\ (\ref{eq:averaged x_n}) 
and  (\ref{eq:xin}) above.
Hence, as long as we are interested in the Lyapunov exponents
only, we can choose freely between the site representation 
(\ref{eq:Wannier})
and the plane wave
representation (\ref{eq:polar decomposition trsf matrix}).

\subsection{Bounded wires: zero energy} 
\label{subsec:Zero energy}

\subsubsection{Wavefunctions}

Zero-modes are solutions to Schr\"odinger equation 
(\ref{eq:matrix H in sublattice grading})
at zero energy. To see how they are constructed, let us first inspect
the case of Fig.\ \ref{fig: 3 exs of lattice with neq BC}(b) in
detail. In this case, the zero
mode is a wavefunction with support on the white sites only.
To construct it, we start with three initial values (seeds) for 
wavefunction on the three leftmost white sites. We now construct
the wavefunction on all other sites in the same way as one finds
the transfer matrix in site representation (\ref{eq:Wannier}):
By applying Schr\"odinger equation to the leftmost black site 
of the middle row one obtains the value of the wavefunction on the 
penultimate leftmost white site of the middle row. Repeating this
process column by column for all black sites that have a white
site to its right, we can construct a wavefunction supported on all 
white sites. For it to be a zero-mode, the Schr\"odinger equation must 
also be satisfied
on the two rightmost black sites that were not used to propagate the 
wavefunction on the white sites. 
For Fig.\ \ref{fig: 3 exs of lattice with neq BC}(b) these are
the two rightmost black sites. Application of Schr\"odinger equation
on each of those sites yields two linear constraints for the
wavefunction elements on the rightmost white sites.
Both constraints can be satisfied since
they are implemented linearly on the three seeds of the 
wavefunction on the left end.

This example suggests a three-step recursive method to obtain zero-modes.
First, the zero-mode has support on one sublattice only.
Second, $N$ independent numbers that make up the $N$--component
vector $\psi_{AL}$ are assigned to the values taken by the wavefunction
on the leftmost sites of each row that belong to sublattice $A$. 
Solution of the Schr\"odinger equation on all sites of sublattice $B$
except for those without a white nearest-neighbor to their left or
right (i.e., except for sites of sublattice $B$ that are at the left 
or right
ends of a row) allows to propagate recursively the wavefunction to
the right column by column.
The rightmost $N$ values of the wavefunction thus constructed 
build the vector $\psi_{AR}$. The relation between the vectors 
$\psi_{AL}$ and $\psi_{AR}$ that is thus obtained can be 
expressed as
\begin{equation}
  \psi_{AR} = M \psi_{AL}.
\label{eq:psiLR}
\end{equation}
The $N\times N$ matrix $M$ is nothing but the counterpart to the
upper left block of the transfer matrix in site representation.
Hence, by Eq.\ (\ref{eq:Mxxs}), it has the polar decomposition
\begin{equation}
  M = 
  U 
  e^X
  V,
  \label{eq:polar decomposition M}
\end{equation}
where $U$ and $V$ are orthogonal (unitary, symplectic) matrices
for $\beta=1$ ($\beta=2,4$), and $X$ is a diagonal matrix with
eigenvalues $x_n$ whose statistics are given by 
Eqs.\ (\ref{eq:averaged x_n}) 
and  (\ref{eq:xin}) above.

Third, in order to have a true zero-mode, the
Schr\"odinger equation must be obeyed on the remaining 
$N_{BL}$ and $N_{BR}$ black sites that do not have white 
nearest-neighbors to their left and right, respectively. This gives $N_{BL}$ 
independent constraints to be satisfied
by the elements of $\psi_{AL}$ and $N_{BR}$ independent constraints 
to be satisfied by the elements of $\psi_{AR}$.
As Eq.\ (\ref{eq:psiLR}) allows for $N$ independent solutions, 
the number of independent zero-modes is equal to 
\begin{eqnarray}
&&
  \mbox{\# zero-modes} = N_A-N_B=N-\left(N_{BL}+N_{BR}\right).
\nonumber \\&&
\label{eq: criterion for existence zero modes}
\end{eqnarray}
The equality $N_A-N_B=N-\left(N_{BL}+N_{BR}\right)$ followed
since the lattice topology is such that
the only black sites without left or right white nearest-neighbors
are located at the far left and far right ends of the wire, respectively.
We emphasize that the criterion 
(\ref{eq: criterion for existence zero modes})
for the {\it existence} of zero-modes does not rely
on the quasi-one-dimensional assumption $N\ll L$ or on the assumption
that there be only one ``transverse'' direction.
The only ingredients needed for 
Eq.\ (\ref{eq: criterion for existence zero modes})
to hold are the boundary conditions at the end of
the disordered region, i.e., the topology of the ``wire''.
The simplifying feature brought by the quasi-one-dimensional limit
$N/L\ll1$,
is that the statistical properties of the transfer matrix
$M$ in Eq.\ (\ref{eq:psiLR}) are known. We will exploit this
knowledge below.

How do the wavefunction elements at the left and right ends of the
wire compare? To answer this question, we first look at the 
geometric mean $\phi$ of $\psi_{AL}$ and $\psi_{AR}$, 
\begin{equation}
  \phi = 
  e^{+X/2}\, 
  V \psi_{AL} = 
  e^{-X/2}\, 
  U^{\dagger} \psi_{AR}.
  \label{eq:phidef}
\end{equation}
In terms of this geometric mean,
the $N_{BL}$ constraints on $\psi_{AL}$ and 
the $N_{BR}$ constraints on $\psi_{AR}$  
can be written in the form
\begin{eqnarray}
  \left( 
  \begin{array}{cc} 
  C_{BL}\, e^{-X/2} 
  \\ 
  C_{BR}\, e^{+X/2} 
  \end{array}
  \right) \phi = 0,
\label{eq:phi}
\end{eqnarray}
where $C_{BL}$ and $C_{BR}$ are $(N_{BL} \times N)$ and $(N_{BR} \times N)$
matrices with coefficients of order unity, respectively. In the
localized regime $L \gg N l$, the $x_n$ are spaced by an amount of
order $L/N l \gg 1$, so that the
coefficients in Eq.\ (\ref{eq:phi}) differ considerably in
magnitude. To see what simplifications this brings about,
we look at the first row of Eq.\ (\ref{eq:phi}),
\begin{eqnarray}
&&
(C_{BL})_{11} e^{-x_1/2}\phi_1
+
\ldots
+
(C_{BL})_{1N} e^{-x_N/2}\phi_N=0. 
\nonumber \\&&
\label{eq:phiN}
\end{eqnarray}
According to Eqs.\ (\ref{eq:averaged x_n},
\ref{eq:xin}), the coefficient 
$(C_{BL})_{1n}e^{-x_n/2}$ is a random number that
fluctuates around $e^{-L/(2\xi_n)}$.
Since by Eq.\ (\ref{eq:xin}) $x_N$ is smaller than all other
$x_n$ by an amount of at least $L/N l \gg 1$, we thus find that 
the left hand side of Eq.\ (\ref{eq:phiN}) is dominated by the 
last term, so that
we conclude $\phi_N = 0$, with exponential accuracy.
Extending this argument to the first $N_{BL}$ and the last $N_{BR}$
rows of Eq.\ (\ref{eq:phi}) we infer that 
\begin{eqnarray}
&&
\phi_n=0,\qquad n=N-N_{BL}+1,\cdots,N,
\nonumber\\
&&
\phi_n=0,\qquad n=1,\cdots,N_{BR},
\end{eqnarray}
respectively, again to exponential accuracy. Conversely, 
the only nonzero elements of $\phi$ are $\phi_n$ with
$n=N_{BR}+1,\ldots,N-N_{BL}$, 
to exponential accuracy, so that, by Eq.\ (\ref{eq:phidef}), 
the only localization lengths available 
to the zero-modes are $\xi_n$ with 
$n=N_{BR}+1,\ldots,N - N_{BL}$.\cite{foot}
In our notations, a negative localization length corresponds to a 
wavefunction exponentially localized near the left end of the wire 
(since then $|\psi_{AL}| \gg |\psi_{AR}|$ in that case), while a positive 
localization lengths correspond to a
wavefunction exponentially localized near the right end of the wire. A
divergent localization length (which can occur for odd $N$)
signals a zero-mode that is critical ($|\psi_{AL}|$ and $|\psi_{AR}|$
comparable in magnitude to exponential accuracy).

We have verified this scenario by numerical implementation of above 
recursive construction of zero-modes in geometries depicted in
Fig.\ \ref{fig: 3 exs of lattice with neq BC}
for various choices of $N$ and of boundary conditions.\cite{Racine}
The agreement found is excellent.

\begin{figure}

\includegraphics[width=\columnwidth]{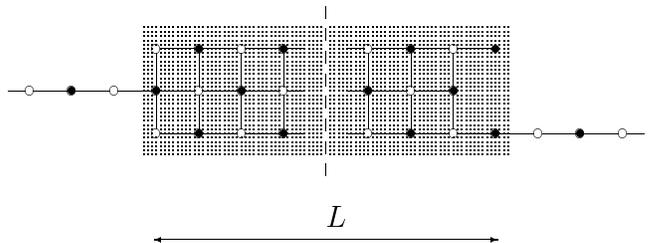}

\caption{\label{fig: wire coupled through point contacts to leads} 
Example of a disordered quantum wire of even length $L$
(dotted region) coupled to left and right reservoirs via point contacts.
In this example the number of left end points on sublattice $A$ is 
$N_{AL}=2$ and the number of right end points on sublattice $B$ is 
$N_{BR}=2$. The number of right end points $N_{AR}$ on sublattice $A$ vanishes
as does the number of left end points $N_{BL}$ on sublattice $B$.
If $L$ is chosen to be odd, $N_{AL}=N_{AR}=2$ whereas
$N_{BL}=N_{BR}=0$. When $L$ is odd, a zero-mode can only be supported on
sublattice $A$ and the conductance must vanish identically.}

\end{figure}

\begin{figure}

\includegraphics[width=\columnwidth]{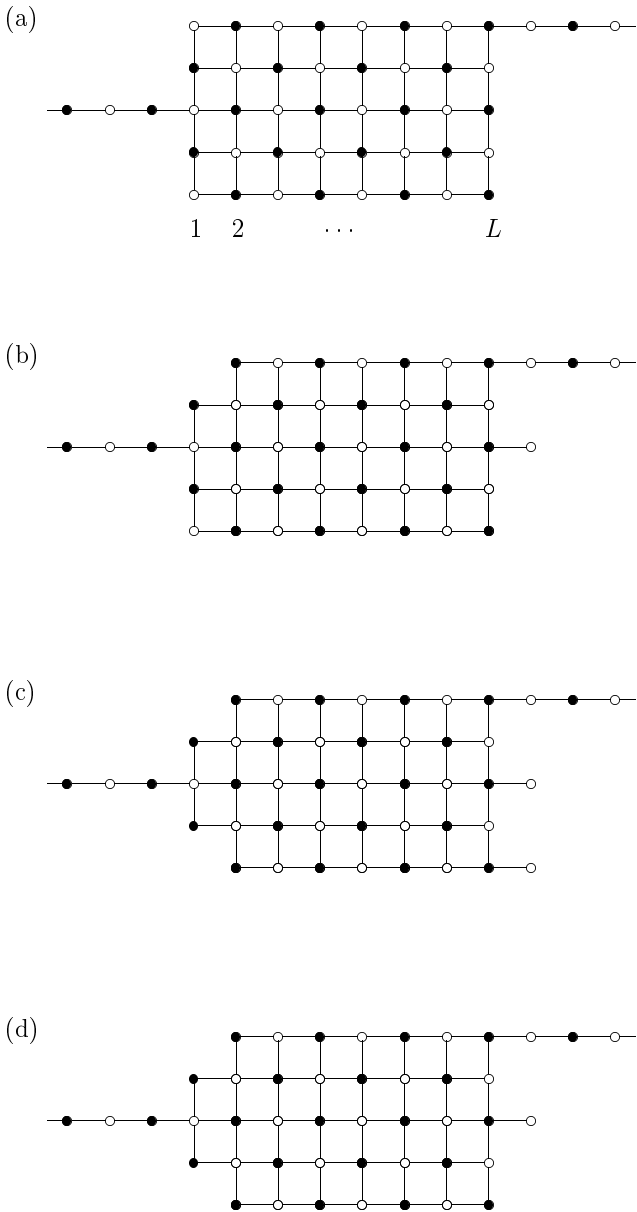}

\caption{\label{fig: geometries for num with N=5}
Four different bounded wires $N=5$ sites wide and $L$ sites long.
In all cases $L$ is  chosen even and the wire is connected to the
reservoirs by single-channel leads. The boundary conditions are
specified by:
(a) $N_{AL}=N_{AR}=N_{BL}=N_{BR}=2$.
(b) $N_{AL}=N_{BR}=1$, $N_{BL}=N_{AR}=3$.
(c) $N_{AL}=N_{BR}=0$, $N_{BL}=N_{AR}=4$.
(d) $N_{AL}=0$, $N_{AR}=3$, $N_{BL}=4$, $N_{BR}=1$.}
\end{figure}

\begin{figure}

\includegraphics[width=\columnwidth]{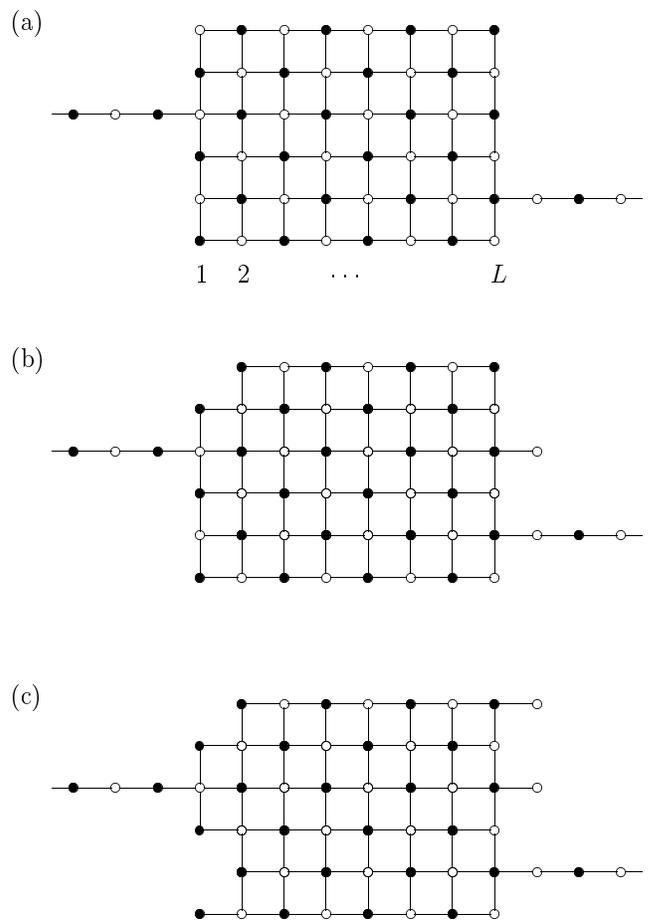}

\caption{\label{fig: geometries for num with N=6}
Three different bounded wires $N=6$ sites wide and $L$ sites long.
In all cases $L$ is  chosen even and the wire is connected to the
reservoirs by single-channel leads. The boundary conditions are
specified by:
(a) $N_{AL}=N_{BR}=2$, $N_{BL}=N_{AR}=3$.
(b) $N_{AL}=N_{BR}=1$, $N_{BL}=N_{AR}=4$.
(c) $N_{AL}=N_{BR}=0$, $N_{BL}=N_{AR}=5$.}
\end{figure}

\subsubsection{Transmission probability}

A different method to probe the effect of boundary conditions
on the localization length of the
random hopping model is via the transmission probability of
a lattice coupled to electron reservoirs via ideal leads. 
Boundary conditions play a role once the width of the leads
is {\it smaller} than the width of the sample lattice, as is shown,
e.g., in 
Figs.\ \ref{fig: wire coupled through point contacts to leads},
       \ref{fig: geometries for num with N=5}
and    \ref{fig: geometries for num with N=6}.

With ideal leads attached to the
left and right ends, zero-modes with support on 
sublattice $A$ can coexist with zero-modes with support on 
sublattice $B$.
In fact, since a traveling wave at zero energy has support on both 
sublattices, conductance through the sample is only possible
if both types of zero-modes exist.\cite{Altland}
Using the same arguments as for the zero-modes in a closed system, 
the possible localization lengths for zero-modes in the
presence of leads can be found from counting the number of 
end points belonging to each sublattice. More precisely,
let $N_{BL}$ and $N_{BR}$ be the number of sites of sublattice $B$
without a white nearest-neighbor to their left and right, respectively.
For all geometries under consideration, $N_{BL}$ ($N_{BR}$)
match the number of end points on the left (right) end of the wire
that belong to sublattice $B$
(sites connected to leads are excluded here).
Now, the quasi-one-dimensional localization lengths of the zero-modes 
with support on 
sublattice $A$ are $\xi_{N_{BR}+1},\ldots,\xi_{N-N_{BL}}$. 
Similarly, the available quasi-one-dimensional 
localization lengths for zero-modes with
support on sublattice $B$ are $\xi_{N_{AR}+1},\ldots,\xi_{N-N_{AL}}$,
where $N_{AL}$ ($N_{AR}$) denotes the numbers of end points on the 
left (right) end of the wire that belong to sublattice $A$
(again, sites connected to leads are here excluded).
Whether or not the quasi-one-dimensional limit applies,
if $N_{BL} + N_{BR} \ge N$ or $N_{AL} + N_{AR} \ge N$, there are 
no zero-modes with support on sublattice $A$ or $B$, respectively,
and hence no traveling waves and no conductance through the sample.
If both $N_{BL} + N_{BR} < N$ and $N_{AL} + N_{AR} < N$, there
is a finite conductance through the sample. In that case, the
quasi-one-dimensional exponential decay length of the conductance,
denoted $\xi^*$,
is the shorter one of the maximum of the decay lengths for zero-modes 
on the two sublattices. We give in 
Tables\ \ref{tab: data output for N=5 geometry}
and
        \ref{tab: data output for N=6 geometry} 
the values taken by $\xi^*$ for the four geometries of
Fig.\ \ref{fig: geometries for num with N=5}
and the three geometries of 
Fig.\ \ref{fig: geometries for num with N=6},
respectively.

\begin{table}
\caption{\label{tab: data output for N=5 geometry}
Maximum localization lengths 
$\xi^*_{A}$ and $\xi^*_{B}$, 
of zero-modes on sublattices $A$ and $B$, respectively, 
for the four different geometries depicted in
Fig.\ \protect\ref{fig: geometries for num with N=5}.
The minimum of
$\xi^*_{A}$ and $\xi^*_{B}$, 
denoted $\xi^*_{\ }$,
controls the conductance.
The entry $\xi^*_{A}=0$ for geometry (d) indicates that no zero-mode
is supported on sublattice A. Correspondingly, the conductance
vanishes at zero energy in this geometry as is implied by the
vanishing localization length $\xi^*_{\ }=0$.}
\begin{ruledtabular}
\begin{tabular}{cccccccc}
Fig.\ \protect\ref{fig: geometries for num with N=5} &
$N_{AL}$                                             &
$N_{AR}$                                             &
$N_{BL}$                                             &
$N_{BR}$                                             &
$\xi^*_A$                                            &
$\xi^*_B$                                            &
$\xi^*_{\vphantom{A}}$                               \\
\hline
(a)                                                  &
2                                                    &
2                                                    &
2                                                    &
2                                                    &
$ \xi^{\vphantom{*}}_3 $                             &
$ \xi^{\vphantom{*}}_3 $                             &
$\infty$                                             \\
(b)                                                  &
1                                                    &
3                                                    &
3                                                    &
1                                                    &
$ \xi^{\vphantom{*}}_2 $                             &
$|\xi^{\vphantom{*}}_4|$                             &
$     {6 l}/{2}$                                     \\
(c)                                                  &
0                                                    &
4                                                    &
4                                                    &
0                                                    &
$ \xi^{\vphantom{*}}_1 $                             &
$|\xi^{\vphantom{*}}_5|$                             &
$     {6 l}/{4}$                                     \\
(d)                                                  &
0                                                    &
3                                                    &
4                                                    &
1                                                    &
0                                                    &
$|\xi^{\vphantom{*}}_4|$                             &
0                                                    \\
\end{tabular}
\end{ruledtabular}

\end{table}

\begin{table}
\caption{\label{tab: data output for N=6 geometry}
Maximum localization lengths 
$\xi^*_{A}$ and $\xi^*_{B}$, 
of zero-modes on sublattices $A$ and $B$, respectively, 
for the three different geometries depicted in
Fig.\ \protect\ref{fig: geometries for num with N=6}.
The minimum of
$\xi^*_{A}$ and $\xi^*_{B}$, 
denoted $\xi^*_{\ }$,
controls the conductance.}
\begin{ruledtabular}
\begin{tabular}{cccccccc}
Fig.\ \protect\ref{fig: geometries for num with N=6} &
$N_{AL}$                                             &
$N_{AR}$                                             &
$N_{BL}$                                             &
$N_{BR}$                                             &
$\xi^*_A$                                            &
$\xi^*_B$                                            &
$\xi^*_{\vphantom{A}}$                               \\
\hline
(a)                                                  &
2                                                    &
3                                                    &
3                                                    &
2                                                    &
$ \xi^{\vphantom{*}}_3 $                             &
$|\xi^{\vphantom{*}}_4|$                             &
$     {7 l}/{1}$                                     \\
(b)                                                  &
1                                                    &
4                                                    &
4                                                    &
1                                                    &
$ \xi^{\vphantom{*}}_2 $                             &
$|\xi^{\vphantom{*}}_5|$                             &
$     {7 l}/{3}$                                     \\
(c)                                                  &
0                                                    &
5                                                    &
5                                                    &
0                                                    &
$ \xi^{\vphantom{*}}_1 $                             &
$|\xi^{\vphantom{*}}_6|$                             &
$     {7 l}/{5}$                                     \\
\end{tabular}
\end{ruledtabular}

\end{table}

Thus, we find that the same range of 
localization lengths shows up in the exponential decay of
wavefunctions in a closed system and of the 
conductance $G$ in an open system, when the wire is coupled to 
the electron reservoirs via point contacts.
This is in stark contrast to the case of 
an ``infinite'' quantum wire 
(i.e., a wire without point contacts at both ends as depicted in
Fig.\ \ref{fig: conventional wire coupled to leads}), where only
the largest localization length determines the conductance. It is also 
in contrast to the case of a quantum wire with standard diagonal disorder, 
where the boundary conditions have no effect
on the exponential decay length of the conductance.

Again, we have verified this scenario and found excellent
agreement between numerics and theoretical expectations.

\begin{figure*}
\centering

\includegraphics[width=88mm]{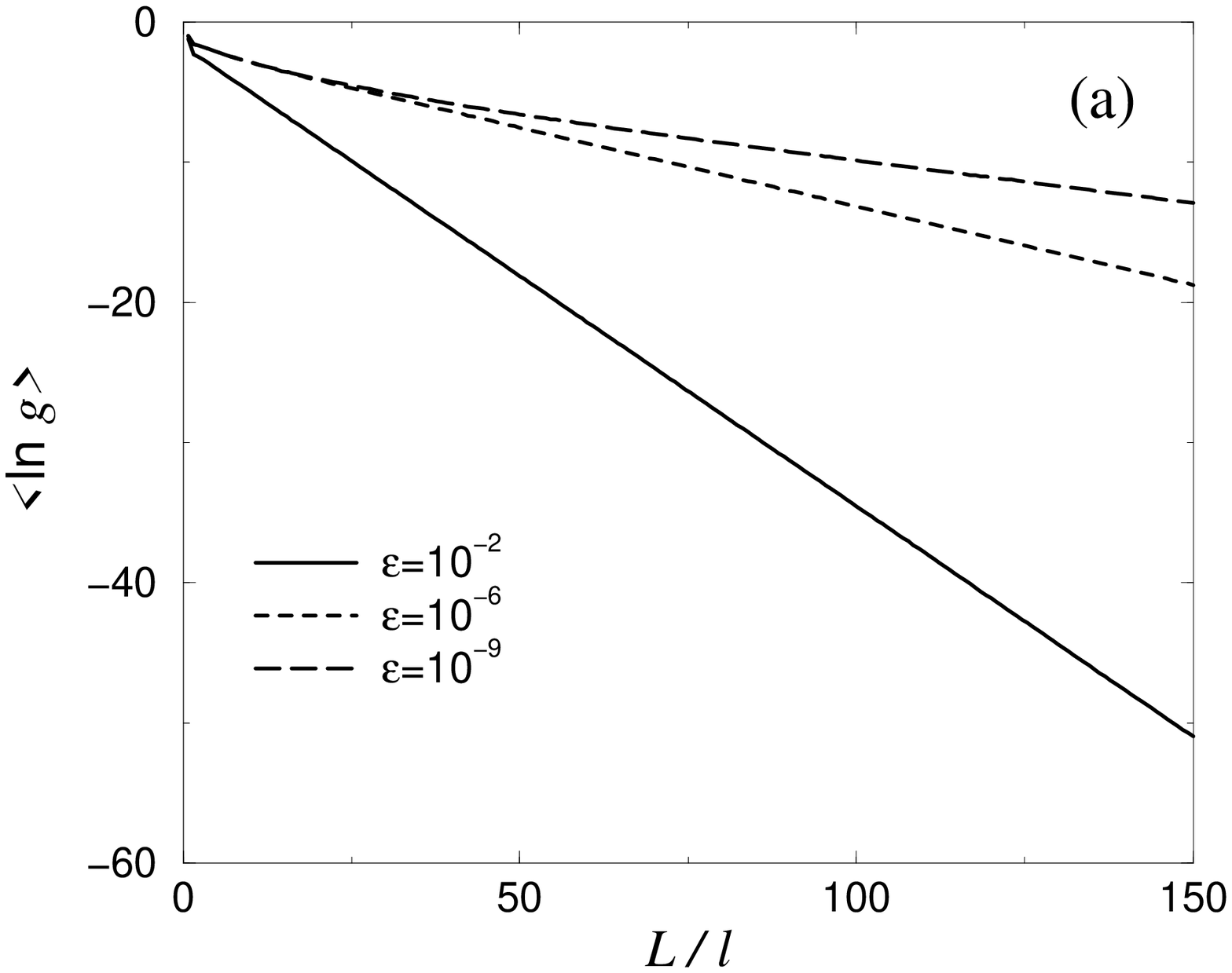}
\includegraphics[width=88mm]{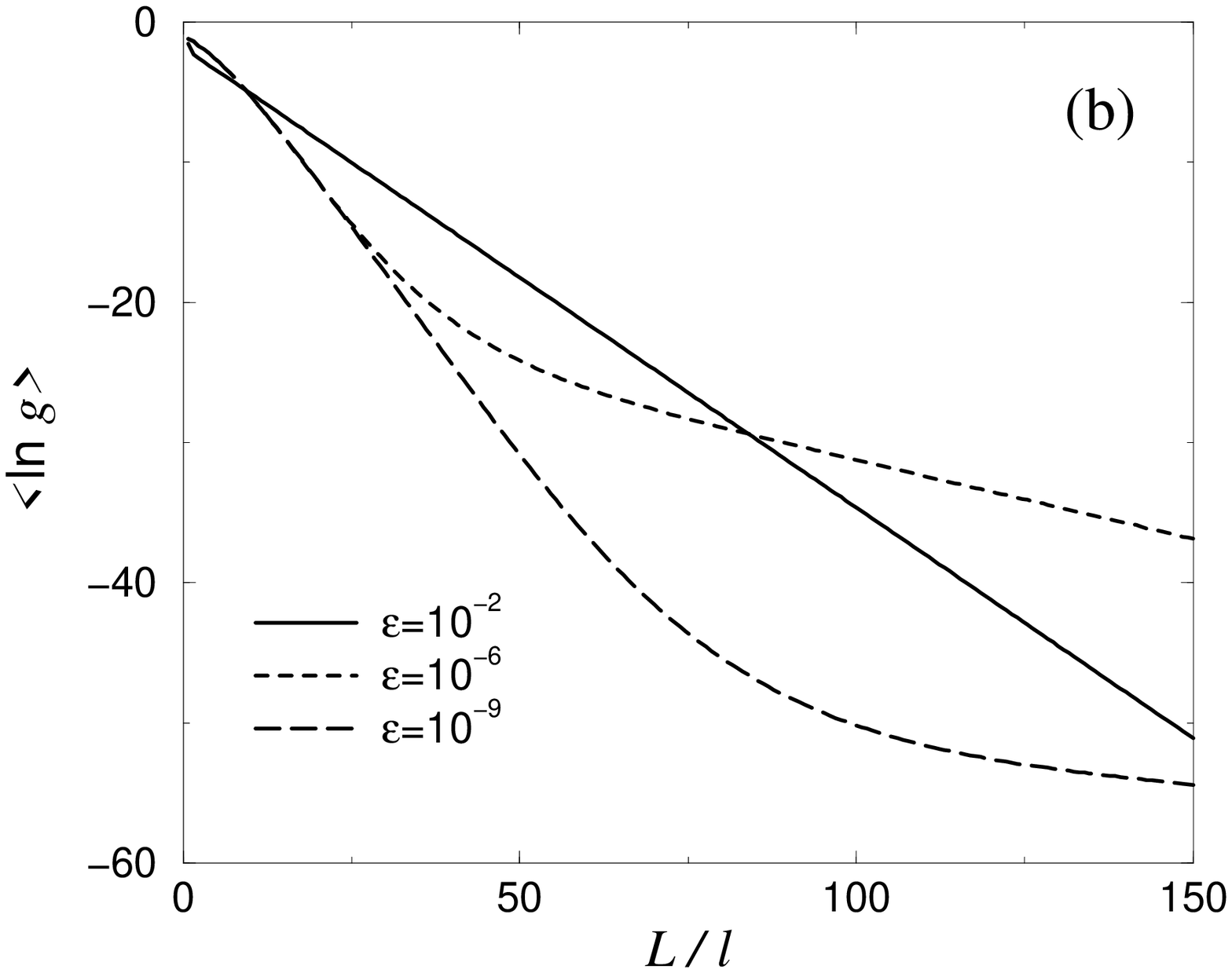}

\includegraphics[width=88mm]{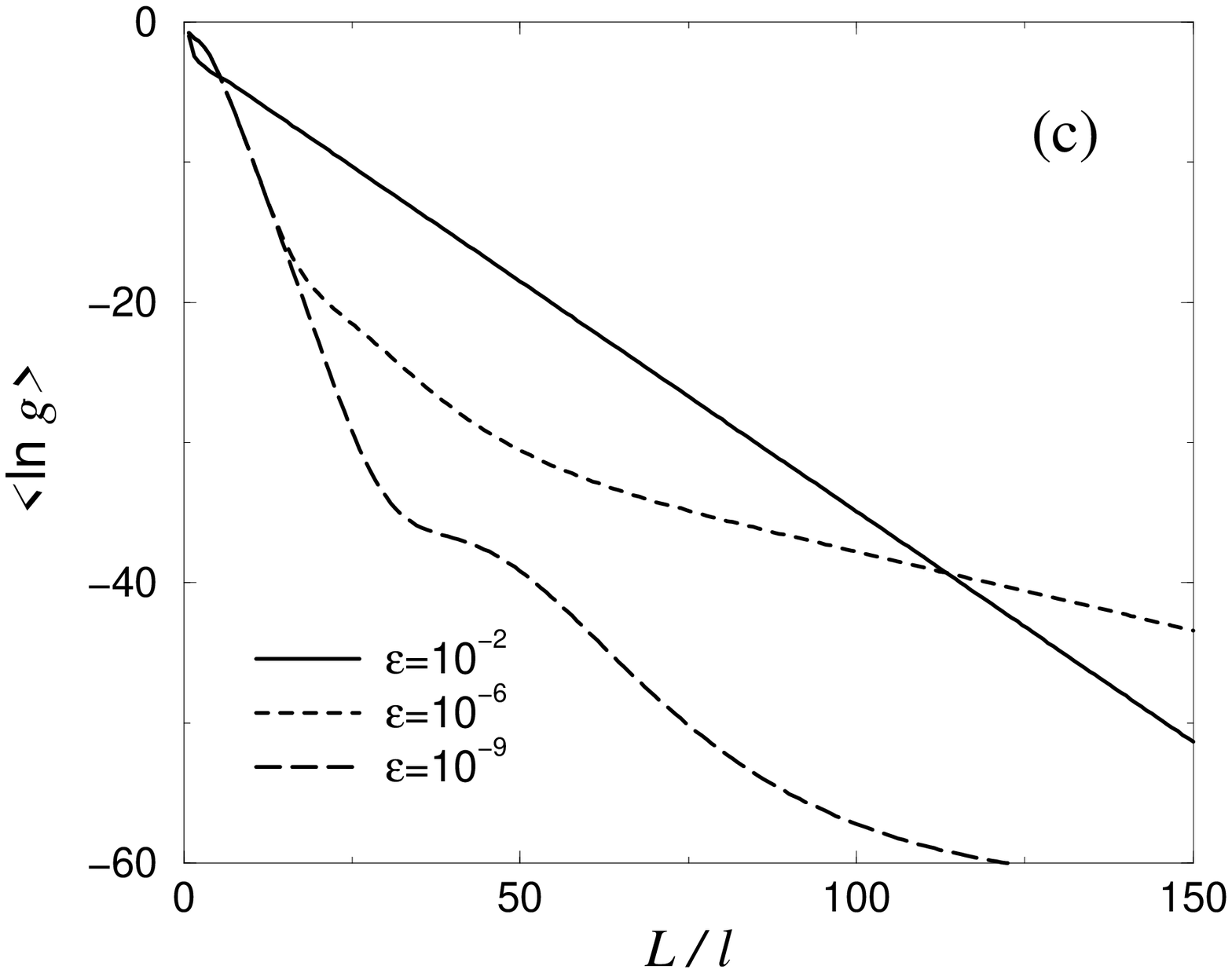}
\includegraphics[width=88mm]{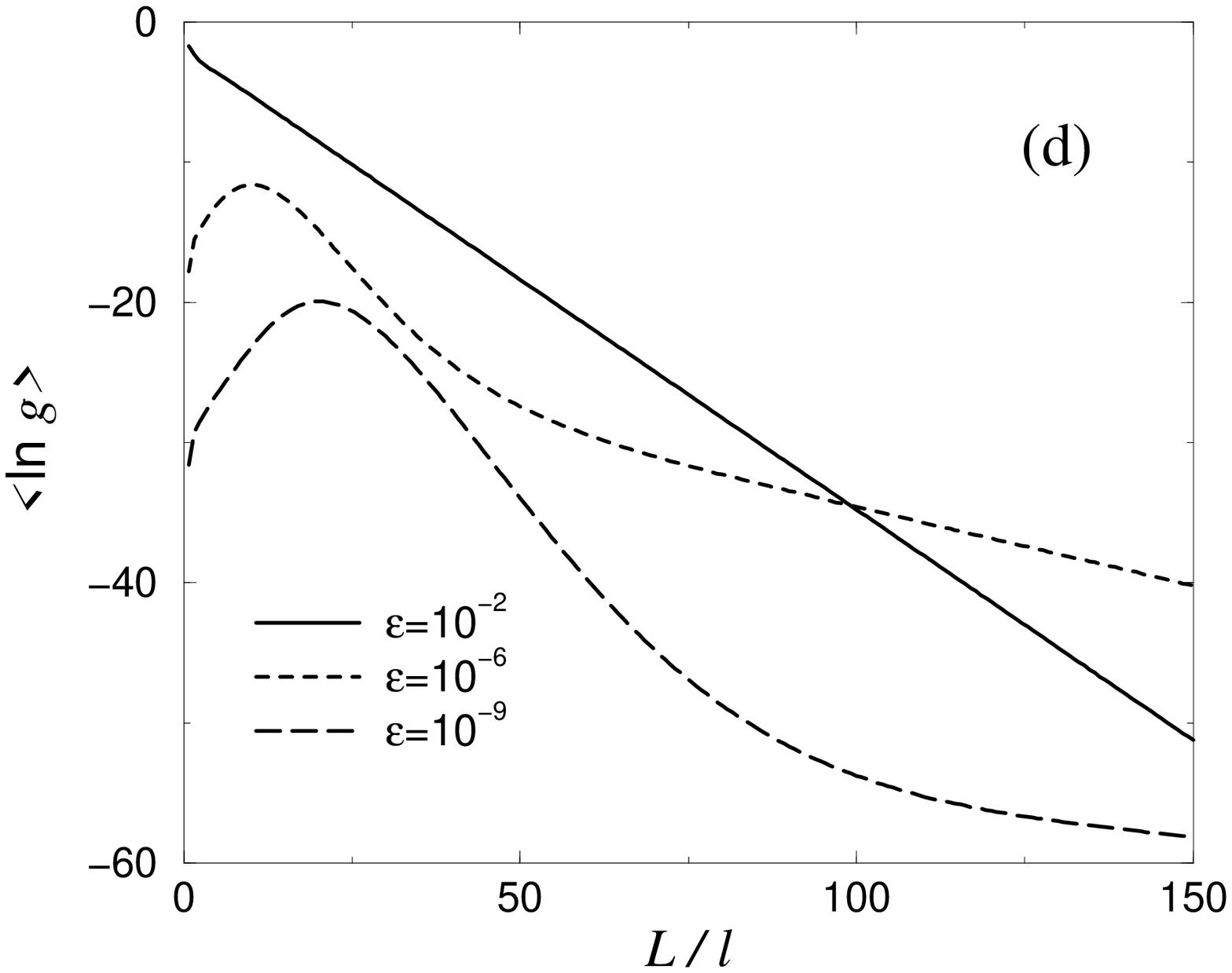}

\caption{\label{fig: Typical conductance N=5}
Dependence on the wire length $L$, $L$ always even, 
of  $\langle \ln g \rangle$
for a quantum wire with random hopping only,
connected to the reservoirs via single-channel ideal leads,
and with the boundary conditions specified in 
Fig.\ \ref{fig: geometries for num with N=5}. 
The quantum wire is $N=5$ sites wide.
The curves represent different values of the energy $\varepsilon$
and different choices for the boundary conditions at the left
and right ends of the wire, as depicted in 
Fig.\ \ref{fig: geometries for num with N=5}.}
\end{figure*}

\begin{figure}

\includegraphics[width=\columnwidth]{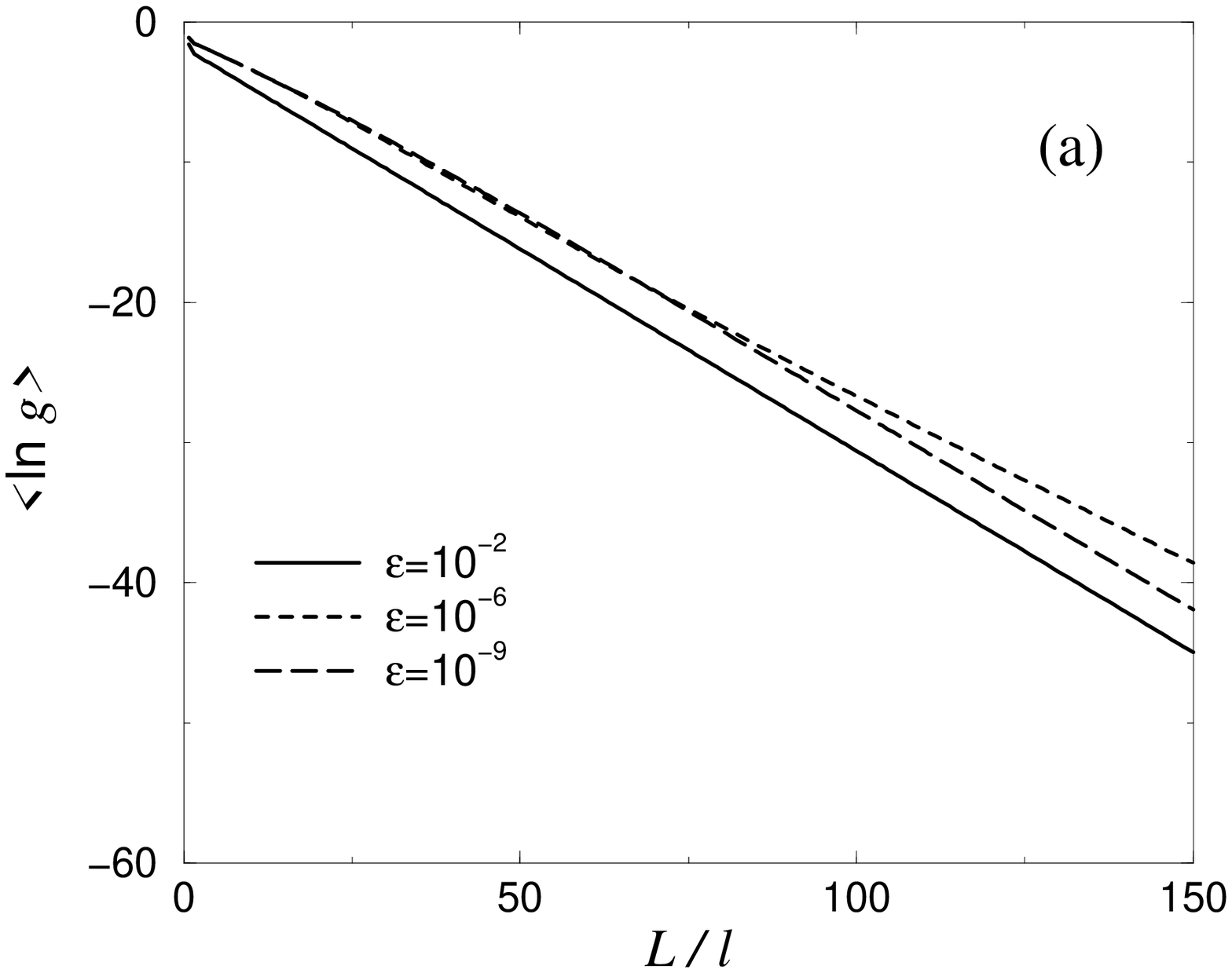}
\vskip -10mm
\includegraphics[width=\columnwidth]{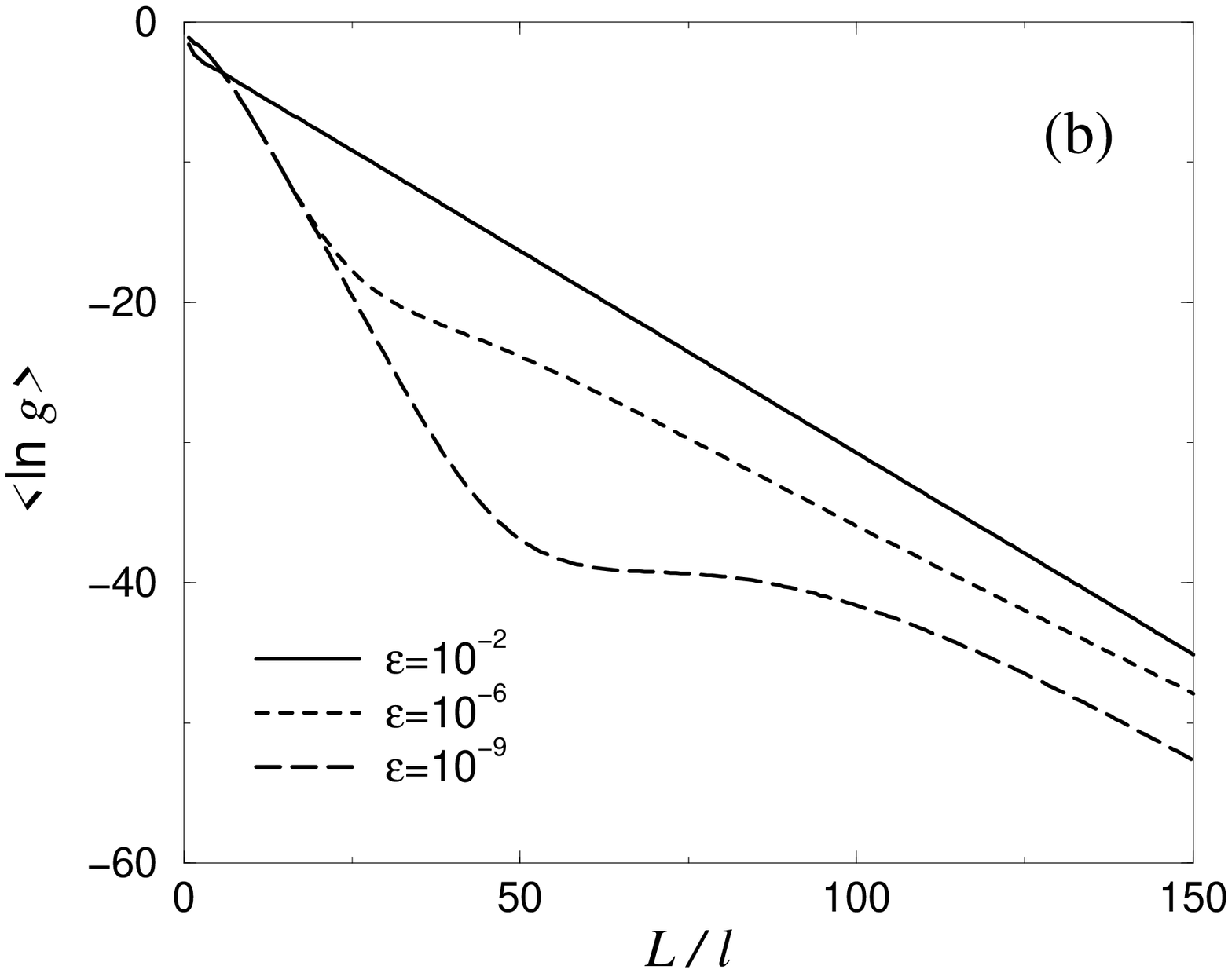}
\vskip -10mm
\includegraphics[width=\columnwidth]{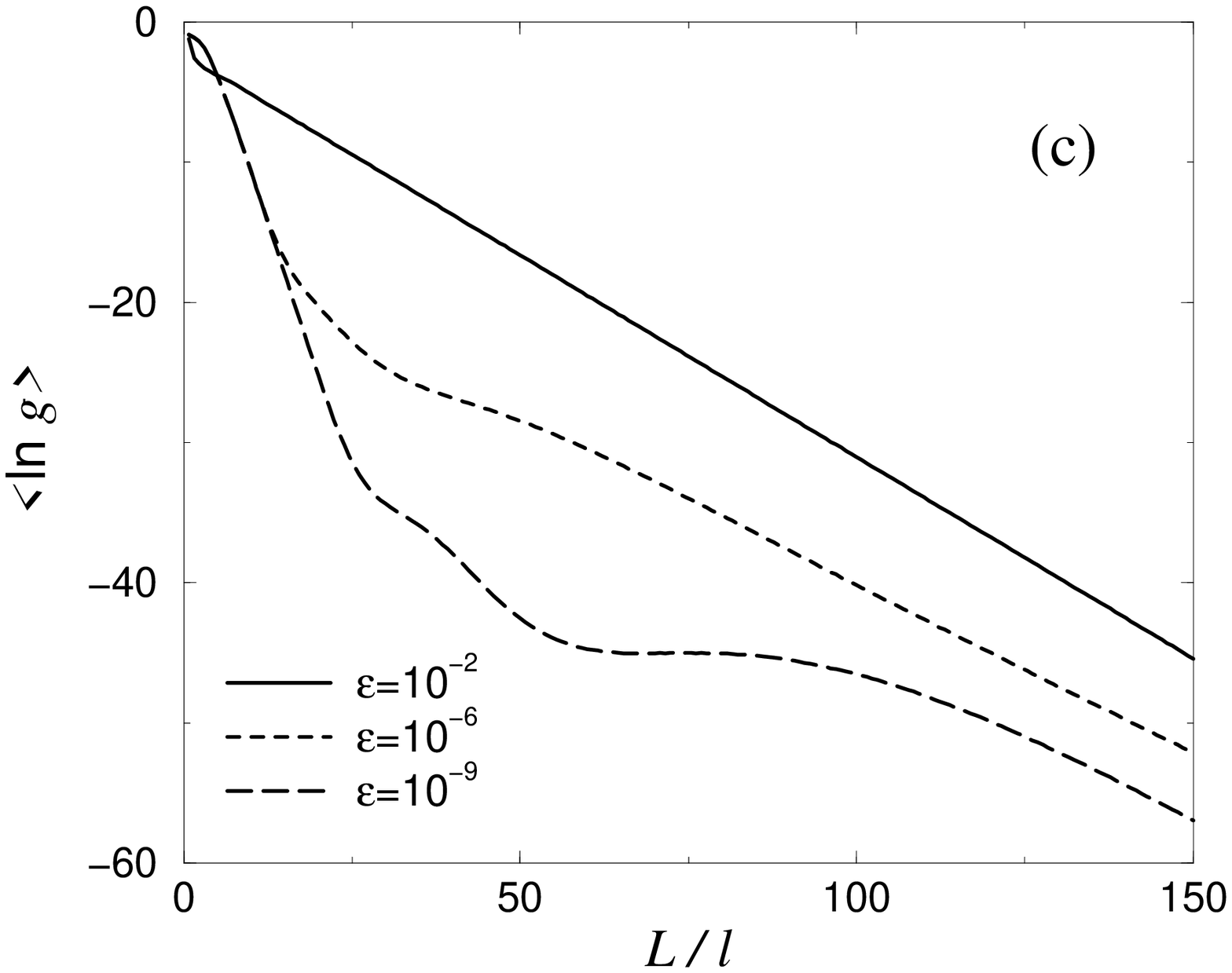}

\caption{\label{fig: Typical conductance N=6}
Dependence on the wire length $L$, $L$ always even, 
of  $\langle \ln g \rangle$
for a quantum wire with random hopping only,
connected to the reservoirs via single-channel ideal leads,
and with the boundary conditions specified in 
Fig.\ \ref{fig: geometries for num with N=6}. 
The quantum wire is $N=6$ sites wide.
The curves represent different values of the energy $\varepsilon$
and different choices for the boundary conditions at the left
and right ends of the wire as depicted in 
Fig.\ \ref{fig: geometries for num with N=6}.}
\end{figure}

\subsection{Bounded wires: nonzero energy} 
\label{subsec:Nonzero energy} 

To study the importance of boundary conditions at finite energy,
we have calculated numerically the conductance of a lattice with 
random hopping
amplitudes and point contacts as a function of energy
using the recursive Green function method.\cite{MacKinnon83,Baranger}
In our numerical simulations we chose real-valued nearest-neighbor
hopping amplitudes $t_{ij}$ in the disordered region 
uniformly and independently
in the intervals $1-\delta<t_{ij}<1+\delta$ for hopping in the
longitudinal direction and $t(1-\delta)<t_{ij}<t(1+\delta)$ in the
transverse direction, with $\delta=0.2$ and $t=0.6$.
With this choice the mean free path $l$ is about 65 lattice
spacings, as can be estimated from a fit to $\langle \ln g \rangle$
versus $L$ for large $L$ and large energy $\varepsilon$ ($\approx
10^{-2}$).
In the perfect leads we take $\delta=0$ and $t=0$.
The numerical data are obtained after averaging over
$10^5$ samples.
The size of error bars for $\langle\ln g\rangle$ is estimated to be
less than 1\%.
A more detailed account of our implementation of the recursive Green 
function method can be found in Refs.\ \onlinecite{Mudry99} and
\onlinecite{Mudry00}.
The disorder strength in the numerical calculations 
presented here is chosen the same as in Ref.\ \onlinecite{Mudry00},
so that a comparison of the results is possible.

\begin{figure}

\includegraphics[width=\columnwidth]{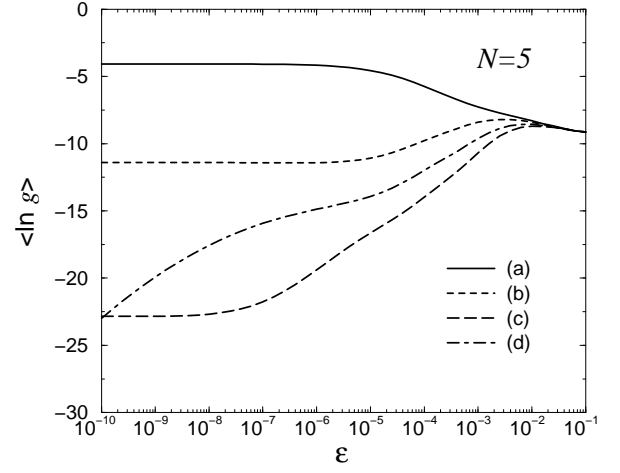}
\includegraphics[width=\columnwidth]{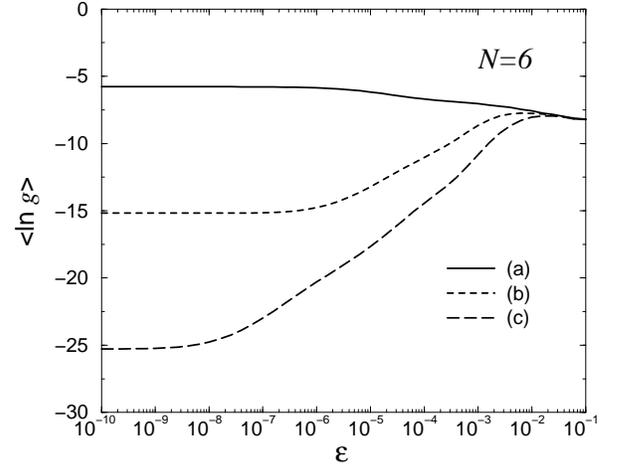}
\caption{\label{fig: crossover as fct energy}
Crossover as a function of energy for a fixed length $L=20 l$ of the
wire of $\langle \ln g \rangle$ for $N=5$ (upper panel) and $N=6$
(lower panel). All traces  saturate to
different \textit{finite} values at very low energies except for trace
(d) of the upper panel for which the conductance vanishes at the
band center.} 
\end{figure}

We have calculated $\langle \ln g \rangle$ versus $L$ for
four different boundary conditions for a quantum wire 
of width $N=5$ and three different boundary conditions for a quantum
wire of width $N=6$. The boundary conditions are shown in 
Figs.\ \ref{fig: geometries for num with N=5}
and    \ref{fig: geometries for num with N=6}. 
Our results are shown in 
Figs.\ \ref{fig: Typical conductance N=5}, 
       \ref{fig: Typical conductance N=6}, 
       \ref{fig: rescaled Lyapunov exponents for N=5,6}, 
and    \ref{fig: crossover as fct energy}. 

Upon increasing the energy away from the band center $\varepsilon=0$, 
a crossover from the chiral (orthogonal) symmetry
class to the standard (orthogonal) symmetry class is expected to 
take place. For wires without point contacts at the left and right
ends, this crossover was studied by three of the authors in Ref.\
\onlinecite{Mudry00}. There, we found that the crossover to the
standard orthogonal class happens for $\varepsilon \sim \varepsilon_c$,
where
\begin{equation}
  \varepsilon_c = \frac{\hbar v_F}{N^2 l}
\end{equation}
is the Thouless energy 
for a localization volume of linear dimension $\sim Nl$.
(The relevant localization length is the smaller one of the
localization lengths in the chiral and standard symmetry
classes). For our calculations we estimate that, with $v_F\approx 2$, 
$\varepsilon_c\approx 10^{-3}$ for $N=5$ and $N=6$.
The largest energy we consider in our calculations, 
$\varepsilon = 10^{-2}$, is well inside the standard orthogonal class, 
see Fig.\ \ref{fig: crossover as fct energy}.
For that largest energy, the four or three curves of 
$\langle \ln g \rangle$ versus $L/l$ 
in 
Figs.\ \ref{fig: Typical conductance N=5} 
and    \ref{fig: Typical conductance N=6}, respectively, 
that correspond to the different boundary conditions are indistinguishable. 
The same conclusion can be reached from 
Figs.\ \ref{fig: rescaled Lyapunov exponents for N=5,6}, 
from which one can infer the Lyapunov exponents 
\begin{eqnarray}
\frac{1}{\xi}=
-\frac{1}{2}\frac{d\langle \ln g \rangle}{dL},
\qquad
L\gg Nl,
\end{eqnarray} 
or from Figs.\ \ref{fig: crossover as fct energy}, 
where we showed the energy dependence
of $\langle \ln g \rangle$ at a fixed length.
This agrees with the
conventional understanding that, in the absence of the
sublattice symmetry, the localization length is
an intrinsic property of the wire, and hence 
boundary-independent.

\begin{figure}

\includegraphics[width=\columnwidth]{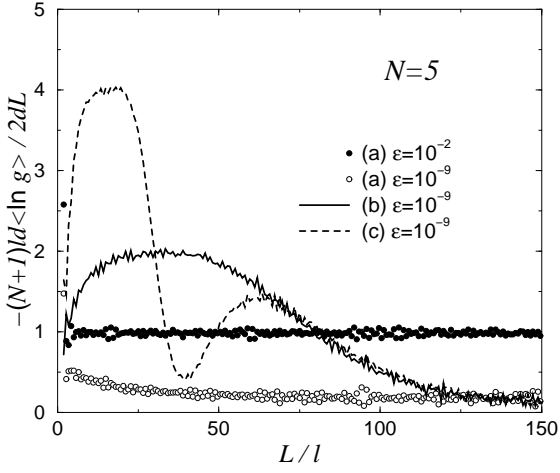}
\includegraphics[width=\columnwidth]{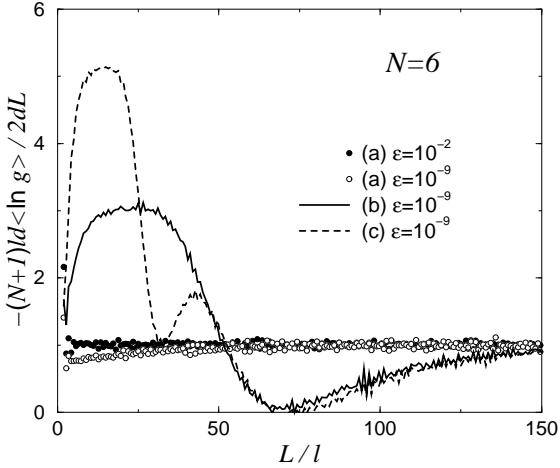}

\caption{\label{fig: rescaled Lyapunov exponents for N=5,6}
Lyapunov exponents $d\langle \ln g \rangle/dL$ for the curves
shown in Figs.\ \protect\ref{fig: Typical conductance N=5} (upper panel)
and \protect\ref{fig: Typical conductance N=6} (lower panel).} 
\end{figure}

The two other energies we considered ($\varepsilon=10^{-6}$ and
$10^{-9}$) are both much smaller than $\varepsilon_c$, 
i.e., well inside the chiral class. 
For short lengths, we see the same \textit{strong} 
dependence on boundary conditions that was predicted for the
zero-modes in the previous subsections. 
A quantitative verification of
the predictions is found from 
Fig.\ \ref{fig: rescaled Lyapunov exponents for N=5,6}, 
where the rescaled Lyapunov exponents 
\begin{equation*}
-\frac{(N+1)l}{2}\frac{d\langle \ln g \rangle}{dL}
\end{equation*} 
take even integer values for odd $N$ and odd integer
values for even $N$ for short $L$ (but still $L \gg Nl$), 
in agreement with Eq.\ (\ref{eq:xin}) with $\beta=1$. 
For larger lengths, however, the dependence on the
boundary conditions is lifted, and the Lyapunov exponents
are the same for all boundary conditions considered
as is illustrated 
in Fig.\ \ref{fig: rescaled Lyapunov exponents for N=5,6}
at energy $\varepsilon=10^{-9}$. 
For a sufficiently large length of the wire 
and upon decreasing the energy,
$-(N+1)l d\langle \ln g \rangle/2dL$
approaches 0  for $N=5$ and $1$ for $N=6$, 
irrespective of the boundary conditions.
Again, this is well illustrated by
Fig.\ \ref{fig: rescaled Lyapunov exponents for N=5,6}
at energy $\varepsilon=10^{-9}$.
Alternatively, for $N=5$ (odd $N$) and energy $\varepsilon = 10^{-9}$,
$\langle \ln g \rangle$ in 
Figs.\ \ref{fig: Typical conductance N=5}(b-c) 
is found to depend 
linearly on length for small $L$ with a boundary-condition dependent
slope, while $\langle \ln g \rangle$ has a curvature consistent with
a $(L/Nl)^{1/2}$ dependence, as is appropriate for critical
conductance statistics. For energy $\varepsilon = 10^{-6}$ the
$L$-dependence of $\langle \ln g \rangle$ is linear for large $L$, but with a
localization length that is significantly larger than for 
$\varepsilon = 10^{-2}$. Such an enhanced localization length 
is characteristic of the crossover between the chiral and 
standard classes for a quantum wire without boundaries.\cite{Mudry00}
For $N=6$ (even $N$) and energies $\varepsilon = 10^{-9}$, $10^{-6}$, 
$\langle \ln g \rangle$ in 
Figs.\ \ref{fig: Typical conductance N=6}(a-c)
decreases linearly with length, 
but with different slopes for small and large $L$. 
These slopes correspond to exponential localization controlled by a
boundary-condition dependent zero mode 
and to exponential localization in the chiral
orthogonal class 
(or, strictly speaking, the crossover between the chiral and
standard orthogonal classes) in an infinite wire, respectively.
(Note, however, the large
energy-dependent crossover
lengths and the non-monotonous length dependence of the
Lyapunov exponents at intermediate length scales
in 
Figs.\ 
\ref{fig: Typical conductance N=5}--\ref{fig: crossover as fct energy}.)
Hence, from the numerical calculations we
conclude that for a finite energy, 
the Lyapunov exponents lose their dependence on the boundary
conditions if the wire is sufficiently long. The typical conductance
itself, $\exp(\langle\ln g\rangle)$,
retains a \textit{strong} dependence on the boundary conditions
for sufficiently long wires through its exponential prefactor
as is illustrated by
Figs.\ \ref{fig: Typical conductance N=5} 
and    \ref{fig: Typical conductance N=6}.
However, the slope of $\langle \ln g\rangle$ as a function of $L/Nl$
in the regime $L/Nl\gg1$
is independent of the boundary conditions.

The length scales where the Lyapunov exponents begin to cross over from the
boundary-condition-dependent value characteristic of the zero-modes
to the ``intrinsic'' (smallest) Lyapunov exponent 
can be estimated as the length scales where the Thouless energy
$\varepsilon_{\mathrm{Th}}(L) = g(L)\Delta(L)$ is equal to the
energy $\varepsilon$, with $g(L) \sim \exp(-2L/|\xi_n|)$ the typical 
boundary-dependent dimensionless conductance of the wire 
($n=1,\ldots,N$)
and 
$\Delta(L) =\frac{\hbar v_F}{N L}$ the mean level spacing of a wire
with length $L$. Hence,
\begin{equation}
  L_{\varepsilon,n} \sim 
  |\xi_{n}| 
  \ln\left(\frac{\hbar v_F}{NL_{\varepsilon,n}\varepsilon}\right),
  \qquad n=1,\ldots,N.
  \label{eq:Lcross}
\end{equation}
No useful crossover length can be defined for the boundary 
condition of Fig.\ \ref{fig: geometries for num with N=5}(d), 
where the zero-energy
conductance of the wire is zero
[see Table\ \ref{tab: data output for N=5 geometry}
 and the downturn of the traces with energies $\varepsilon=10^{-6},10^{-9}$
for sufficiently small wire lengths in
     Fig.\ \ref{fig: Typical conductance N=5}(d)].
Equation (\ref{eq:Lcross})
implies that the shorter $|\xi_n|$ disappear at shorter wire
lengths than the larger $|\xi_n|$.
All dependence on boundary conditions is removed and only the smallest
of the Lyapunov exponents survives
for lengths larger than the second largest of the $L_{\varepsilon,n}$, i.e.,
beyond $L_{\varepsilon} \equiv L_{\varepsilon,(N-1)/2}$ for $N$ odd and
beyond $L_{\varepsilon} \equiv L_{\varepsilon,(N-2)/2}$ for $N$ even.
This is well illustrated in 
Figs.\ \ref{fig: rescaled Lyapunov exponents for N=5,6}, where the
shortest localization length (corresponding to the highest Lyapunov
exponent) disappears first, to be followed by the second-shortest
localization length at a wire size that is roughly a factor $2$
($N=5$) or $5/3$ ($N=6$) larger 
(see Tables\ \ref{tab: data output for N=5 geometry}
and          \ref{tab: data output for N=6 geometry},
respectively).

To summarize: While the exponential decay length of the conductance
depends on the boundary conditions for $\varepsilon=0$ even in
the limit $L \to \infty$, the exponential decay of the conductance
is governed by the ``intrinsic'' (largest) localization length $\xi$
for {\em any} finite energy $\varepsilon$ different from zero. In
this sense, the remarkable dependence of the zero-mode localization
lengths and of the zero energy conductance on boundary conditions
can be considered as an anomaly belonging to the case of energy
being exactly equal to zero, not as something representative of the 
thermodynamic limit of the random hopping model.

\begin{figure}

\includegraphics[width=\columnwidth]{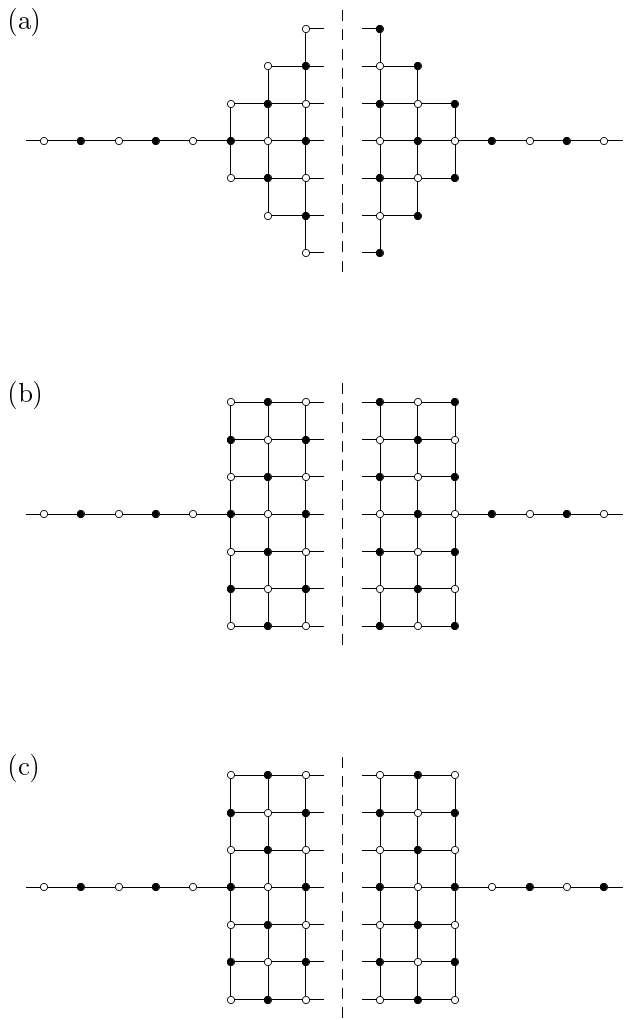}

\caption{\label{fig: lattices in higher dim} 
Three examples of two-dimensional lattices that
have different size-dependencies of the conductance, see text.}
\end{figure}

\section{Higher dimensional examples} 
\label{sec:Higher dimensional examples}

The examples we have discussed so far pertain to a quasi-one-dimensional
geometry. We would like to close with examples of lattices that are
extended in two or more dimensions. For the three lattices shown in 
Fig.\ \ref{fig: lattices in higher dim}, 
three different scenarios apply to the 
zero-energy conductance between the left and right leads. 

For the lattice of 
Fig.\ \ref{fig: lattices in higher dim}(a), 
the conductance decays exponentially with a decay length of the order of the
mean free path of the system as can be seen by direct simulation of the
conductance or by constructing recursively zero-modes on each sublattices. 
In the latter case, a seed value is assigned first to the leftmost (rightmost)
black (white) site having a pair of white (black) nearest-neighbor sites in
the direction orthogonal to the leads. The Schr\"odinger 
equation is then solved
on the white (black) sites moving to the right (left) column by column.
The zero-mode on black sites is exponentially localized near the 
right contact and the zero-mode on white sites is exponentially localized 
near the left contact with a localization length of the order of the
mean free path. The exponentially small conductance at the band center
is found in all ``diamond''-like structures of the form of Fig.\
\ref{fig: lattices in higher dim}(a), 
irrespective of dimensionality and disorder strength.

One possible generalization of the ``diamond''-like geometry of 
Fig.\ \ref{fig: lattices in higher dim}(a) 
is to couple sites on the lower left and upper right edges 
of the diamond to the reservoirs. In this case, the number of zero-modes 
scales with the width of the leads.
To each seed on one edge corresponds a zero-mode exponentially localized about
its vis a vis on the opposite edge. Linear superposition of these zero-modes
yield traveling waves on the edges that are exponentially localized in 
the direction orthogonal to the edges. This is reminiscent of
the Callan-Harvey effect\cite{Callan85} in field theory.

For the ``square'' lattice of 
Fig.\ \ref{fig: lattices in higher dim}(c), 
the conductance is zero, as
this geometry does not allow a zero-mode on the black sites according
to the arguments of Subsec.\ \ref{subsec:Zero energy}.
(Recall that the condition of quasi-one-dimensionality is not needed
 to establish the conditions for the existence of zero-modes.)

Finally, the lattice of 
Fig.\ \ref{fig: lattices in higher dim}(b) 
has zero-modes on white sites and
on black sites that are believed to be critical 
(i.e., not exponentially localized). This is not surprising in
a quasi-one-dimensional geometry for which the 
lattice only extends in the longitudinal direction, as can be 
seen using the arguments of Sec.\ \ref{subsec:Zero energy}.
More surprising is that the critical nature of the zero-mode
seems not to depend on the transversal extension.  
Verg\'es in Ref.\ \onlinecite{Verges01} has studied numerically the 
conductance distribution in the geometry of 
Fig.\ \ref{fig: lattices in higher dim}(b)
for a square lattice made up of up to $799\times799$ sites.
His conclusion is that the probability distribution in the geometry of 
Fig.\ \ref{fig: lattices in higher dim}(b) 
is best fitted by the conductance distribution of a 
thick quantum wire with an odd number $N$ of channels
derived in Ref.\ \onlinecite{Mudry00}.
He thus concludes that the critical zero-mode 
in the two-dimensional geometry of 
Fig.\ \ref{fig: lattices in higher dim}(b)
is quasi-one-dimensional in nature.

\section{conclusions}
\label{sec:Conclusions}

In this paper we have investigated the dependence on boundary
conditions of localization properties of the
random hopping problem at the band center and its vicinity.
At finite energies, localization properties are intrinsic, i.e.,
independent of boundary conditions in the thermodynamic limit.
Remarkably, this is not true anymore precisely at the band center
where both the transmission probability of a disordered region
connected to reservoirs by single-channel leads
and the spatial decay of zero-modes in closed systems
are highly sensitive to the choice of boundary conditions
even as the thermodynamic limit is taken.
This sensitivity to boundary conditions was quantified analytically
in quasi-one-dimensional geometries. In particular,
the conditions under which zero-modes are critical were given. 
In higher dimensions, one must rely on numerical simulations
to  study localization properties of zero-modes. However, 
the task is somehow simplified
by an explicit recursive construction of zero-modes that we showed applies to
a large class of geometries. It is an
interesting open problem to determine conditions for criticality of
zero-modes beyond quasi-one-dimensional geometries
and whether a field theoretical description of intrinsic critical
properties in terms of critical zero-modes
in the spirit of 
Refs.\ \onlinecite{Kogan96},\onlinecite{Shelton98},\onlinecite{Balents97}
and \onlinecite{Hastings01}
applies to the two-dimensional random hopping problem.


\section*{Acknowledgments}

We would like to thank A.\ Altland for discussions.
This work was supported in part 
by the NSF under grant no.\ DMR 0086509 
and by the Sloan and Packard foundations (PWB);
by Fonds pour la Formation de Chercheur et l'Aide \` a la Recherche (ER);
by Grant-in-Aid for Scientific Research on Priority Areas (A) from the
Ministry of Education, Science, Sports and Culture (No. 12046238) (AF).
PWB and AF thank the Institute for Theoretical Physics
in Santa Barbara for its hospitality during the final stages of
this work. CM thanks the Yukawa Institute for Theoretical Physics in Kyoto
for its hospitality during the final stages of this work.
Numerical data presented here were obtained at the Yukawa Institute
Computing facility.


\begin{thebibliography}{ourpaper}

\bibitem{Callan85}
C.\ G.\ Callan and J.\ A.\ Harvey, 
Nucl.\ Phys.\ \textbf{B250}, 427 (1985).

\bibitem{Janssen96}
For a review, see
K.\ Janssen,
Phys.\ Rep.\ \textbf{273}, 1 (1996).
See also 
Refs.\ \onlinecite{Kohmoto89,Su79,Su81,Fradkin86,Weaire71,Fleishman77,Stone84,Sutherland86,Inui94,Hatsugai97}.

\bibitem{Schwinger62}
J.\ Schwinger,
Phys.\ Rev.\ \textbf{128}, 2425 (1962).

\bibitem{Aharonov79}
Y.\ Aharonov and A.\ Casher,
Phys.\ Rev.\ A \textbf{19}, 2461 (1979).

\bibitem{Moroz95}
A.\ Moroz,
Phys.\ Lett.\ B \textbf{358}, 305 (1995).

\bibitem{Kohmoto89}
M.\ Kohmoto,
Phys.\ Rev.\ B, \textbf{39}, 11943 (1989).

\bibitem{Witten85}
E.\ Witten, 
Nucl.\ Phys.\ \textbf{B249}, 557 (1985).

\bibitem{Jackiw76}
R.\ Jackiw and C.\ Rebbi, 
Phys.\ Rev.\ D \textbf{13}, 3398 (1976).

\bibitem{Su79}
W.\ P.\ Su, J.\ R.\ Schrieffer, and A.\ J.\ Heeger,
Phys.\ Rev.\ Lett.\ \textbf{42}, 1698 (1979).

\bibitem{Su81}
W.\ P.\ Su and J.\ R.\ Schrieffer,
Phys.\ Rev.\ Lett.\ \textbf{46}, 738 (1981).

\bibitem{Takayama80}
H.\ Takayama, Y.R.\ Lin-liu, and K.\ Maki,
Phys.\ Rev.\ B \textbf{21}, 2388 (1980),
see also
R.\ Jackiw and J.\ R.\ Schrieffer,
Nucl.\ Phys.\ \textbf{B190}, 253 (1981).

\bibitem{Golstone86}
J.\ Goldstone and F.\ Wilczek, 
Phys.\ Rev.\ Lett.\ \textbf{47}, 986 (1981).

\bibitem{Su86}
Z.-B.\ Su and B.\ Sakita,
Phys.\ Rev.\ Lett.\ \textbf{56}, 780 (1986).

\bibitem{Stone85}
M.\ Stone, A.\ Garg, and P.\ Muzikar,
Phys.\ Rev.\ Lett.\ \textbf{55}, 2328 (1985).

\bibitem{Volovik86}
G.\ E.\ Volovik,
Pis'ma Zh.\ Eksp.\ Teor.\ Fiz.\ \textbf{43}, 428 (1986)
[JETP Lett.\ \textbf{43}, 551 (1986)].

\bibitem{Gaitan87}
M.\ Stone and F.\ Gaitan, 
Ann.\ Phys.\ (N.Y.) \textbf{178}, 89 (1987).

\bibitem{Fradkin86}
E.\ Fradkin, E.\ Dagotto, and D.\ Boyanovsky, 
Phys.\ Rev.\ Lett.\ \textbf{57}, 2967 (1986);
Nucl.\ Phys.\ \textbf{B285}, 340 (1987).

\bibitem{Hu94}
C.-R.\ Hu,
Phys.\ Rev.\ Lett.\ \textbf{72}, 1526 (1994),

\bibitem{Volovik97}
G.\ E.\ Volovik, 
Pis'ma Zh.\ \'Ekps.\ Teor.\ Fiz.\ \textbf{66}, 492 (1997)
[JETP Lett.\ \textbf{66}, 522 (1997)].

\bibitem{Laughlin98}
R.\ B.\ Laughlin, 
Phys.\ Rev.\ Lett.\ \textbf{80}, 5188 (1998).

\bibitem{Buchholtz81}
L.\ J.\ Buchholtz and G.\ Zwicknagl, 
Phys.\ Rev.\ B \textbf{23},  5788 (1981).

\bibitem{Wakabayashi00}
K.\ Wakabayashi, M.\ Fujita, H.\ Ajiki, and M.\ Sigrist, 
Phys.\ Rev.\ B \textbf{59}, 8271 (1999).
%
%

\bibitem{Lieb89}
E.\ H.\ Lieb,
Phys.\ Rev.\ Lett.\ \textbf{62}, 1201 (1989).

\bibitem{Weaire71}
D.\ Weaire and M.\ F.\ Thorpe,
Phys.\ Rev.\ A \textbf{4}, 2508 (1971).

\bibitem{Fleishman77}
L.\ Fleishman and D.\ C.\ Licciardello,
J.\ Phys.\  C \textbf{10}, L125 (1977).

\bibitem{Stone84}
M.\ Stone,
Ann.\ Phys.\ (N.Y.) \textbf{155}, 56 (1984).

\bibitem{Sutherland86}
B.\ Sutherland,
Phys.\ Rev.\ B \textbf{34}, 5208 (1986).

\bibitem{Inui94} 
M.\ Inui, S.\ A.\ Trugman, and E.\ Abrahams, 
Phys.\ Rev.\ B \textbf{49}, 3190 (1994).

\bibitem{Hatsugai97}
Y.\ Hatsugai, X.-G.\ Wen, and M.\ Kohmoto,
Phys.\ Rev.\ B\ \textbf{56}, 1061 (1997);
Y.\ Morita and Y.\ Hatsugai,
Phys.\ Rev.\ Lett.\ \textbf{79}, 3728 (1997);
Phys.\ Rev.\ B\ \textbf{58}, 6680 (1998).

\bibitem{Altland99}
A.\ Altland and B.\ D.\ Simons,
Nucl.\ Phys.\ \textbf{B562}, 445 (1999);
\textit{ibid}, J.\ Phys.\ \textbf{A32}, L353 (1999).

\bibitem{RMT}
D.\ A.\ Ivanov,
J.\ Math.\ Phys.\ \textbf{43}, 126 (2002);
and references therein.

\bibitem{Dyson53}
F.\ J.\ Dyson, 
Phys.\ Rev.\ \textbf{92}, 1331 (1953).

\bibitem{Theodorou76}
G.\ Theodorou and M.\ H.\ Cohen,
Phys.\ Rev.\ B \textbf{13}, 4597 (1976).

\bibitem{Eggarter78}
T.\ P.\ Eggarter and R.\ Riedinger,
Phys.\ Rev.\ B \textbf{18}, 569 (1978).


\bibitem{Shelton98}
D.\ G.\ Shelton and A.\ M.\ Tsvelik,
Phys.\ Rev.\ B \textbf{57}, 14242 (1998).

\bibitem{Hastings01}
M.\ B.\ Hastings and S.\ L.\ Sondhi,
Phys.\ Rev.\ B \textbf{64}, 094204 (2001).

\bibitem{Fisher94}
D.\ S.\ Fisher,
Phys.\ Rev.\ B \textbf{50}, 3799 (1994); 
\textbf{51}, 6411 (1995).

\bibitem{Balents97}
L.\ Balents and  M.\ P.\ A.\ Fisher,
Phys.\ Rev.\ B \textbf{56}, 12970 (1997).

\bibitem{Ludwig94}
A.\ W.\ W.\ Ludwig, M.\ P.\ A.\ Fisher, R.\ Shankar, and G.\ Grinstein,
Phys.\ Rev.\ B \textbf{50}, 7526 (1994).

\bibitem{Nersesyan94}
A.\ A.\ Nersesyan, A.\ M.\ Tsvelik, and F.\ Wenger,
Phys.\ Rev.\ Lett.\ \textbf{72}, 2628 (1994);
Nucl.\ Phys.\ \textbf{B438}, 561 (1995).

\bibitem{Mudry96}
C. Mudry, C.\ Chamon, and X.-G. Wen, 
Nucl.\ Phys.\ \textbf{B466}, 383 (1996).

\bibitem{Chamon96}
C.\ C.\ Chamon, C.\ Mudry, and X.-G. Wen, 
Phys.\ Rev.\ Lett.\ \textbf{77}, 4194 (1996).

\bibitem{Kogan96}
I.\ I.\ Kogan, C.\ Mudry, and A.\ M.\ Tsvelik,
Phys.\ Rev.\ Lett.\  \textbf{77}, 707 (1996).


\bibitem{Brouwer98}
P.\ W.\ Brouwer, C.\ Mudry, B.\ D.\ Simons, and A.\ Altland, 
Phys.\ Rev.\ Lett.\ \textbf{81}, 862 (1998).

\bibitem{Beenakker97}
C.\ W.\ J.\ Beenakker, 
Rev.\ Mod.\ Phys.\ \textbf{69}, 731 (1997).

\bibitem{Brouwer00}
P.\ W.\ Brouwer, C.\ Mudry, and A.\ Furusaki, 
Nucl.\ Phys.\ \textbf{B565}, 653 (2000).

\bibitem{foot}
Strictly speaking, the above argument does not describe the
envelope of the entire wavefunction, but only the relative
magnitudes of the zero-mode at the left and right ends of the
wire. However, since the very existence depends on the boundary
conditions, we can rule out the possibility of a wavefunction that
is localized inside the sample. This conclusion is confirmed by
our numerical simulations.

\bibitem{Racine}
E.\ Racine, unpublished.

\bibitem{Altland} 
A.\ Altland and R.\ Merkt, 
Nucl.\ Phys.\ \textbf{B607}, 511 (2001).

\bibitem{MacKinnon83} 
A.\  MacKinnon and B.\  Kramer, Z.\  Phys.\ B \textbf{53}, 1 (1983). 

\bibitem{Baranger}
H.\  U.\  Baranger, D.\  P.\  DiVincenzo, R.\  A.\  Jalabert, and 
A.\  D.\  Stone, Phys.\ Rev.\ B \textbf{44}, 10 637 (1991).

\bibitem{Mudry99}
C.\ Mudry, P.\ W.\ Brouwer, and A.\ Furusaki, 
Phys.\ Rev.\ B \textbf{59}, 13221 (1999)

\bibitem{Mudry00} 
C.\ Mudry, P.\ W.\ Brouwer, and A.\ Furusaki, 
Phys.\ Rev.\ B \textbf{62}, 8249 (2000); 
Phys.\ Rev.\ B \textbf{63}, 129901 (2001) [E]. 

\bibitem{Verges01}
J.\ A.\ Verg\'es, 
Phys.\ Rev.\ B \textbf{65}, 054201 (2002).


\end{thebibliography}
\end{document}